\documentclass[journal]{IEEEtran}

\usepackage{cite}
\usepackage[numbers,sort&compress]{natbib}
\usepackage{graphicx}
\usepackage{subfigure}
\usepackage{booktabs}
\usepackage{amsfonts,amssymb}
\usepackage{mathrsfs}
\usepackage{algorithm}
\usepackage{algorithmicx}
\usepackage{algpseudocode}
\usepackage{amsmath}
\usepackage{multirow}
\usepackage{colortbl}
\definecolor{mygray}{gray}{.6}
\usepackage{array}

\begin{document}
%
\title{RLTIR: Activity-based Interactive Person Identification via Reinforcement Learning Tree}
%
%
%

\author{Qingyang Li,
        Zhiwen Yu,~\IEEEmembership{Senior Member,~IEEE},\\
        Lina Yao,~\IEEEmembership{Member,~IEEE},~and~Bin Guo,~\IEEEmembership{Senior Member,~IEEE}
\thanks{Q. Li, Z. Yu, and B. Guo are with the School
of Computer Science, Northwestern Polytechnical University, Xi'an 710129, China (e-mail: qingyangli@mail.nwpu.edu.cn, zhiwenyu@nwpu.edu.cn, guob@nwpu.edu.cn). 
L. Yao is with the School of Computer Science and Engineering, University of New South Wales, Kensington, NSW, Australia. (e-mail: lina.yao@unsw.edu.au).}
\thanks{Manuscript received April 19, 2005; revised August 26, 2015.}}

%
%

\markboth{Journal of \LaTeX\ Class Files,~Vol.~14, No.~8, August~2015}%
{Shell \MakeLowercase{\textit{et al.}}: Bare Demo of IEEEtran.cls for IEEE Journals}
%



\maketitle

\begin{abstract}
Identity recognition plays an important role in ensuring security in our daily life. Biometric-based (especially activity-based) approaches are favored due to its fidelity, universality, and resilience. However, most existing machine learning-based approaches rely on a traditional workflow where models are usually trained once for all, with limited involvement from end-users in the process and neglecting the dynamic nature of the learning process. This makes the models static and can not be updated in time, which usually leads to high false positive or false negative. Thus, in practice, an expert is desired to assist with providing high-quality observations and interpretation of model outputs. 
It is expedient to combine both advantages of human experts and the computational capability of computers to create a tight-coupling incremental learning process for better performance. In this study, we develop RLTIR, an interactive identity recognition approach based on reinforcement learning, to adjust the identification model by human's guidance. We first build a base tree-structured identity recognition model. And an expert is introduced in the model for giving feedback upon model outputs. Then, the model is updated according to strategies that are automatically learned under a designated reinforcement learning framework. To the best of our knowledge, it is the very first attempt to combine human expert knowledge with model learning in the area of identity recognition. The experimental results show that the reinforced interactive identity recognition framework outperforms baseline methods with regards to recognition accuracy and robustness.
\end{abstract}


%
\IEEEpeerreviewmaketitle

\section{Introduction}
\IEEEPARstart{P}{erson} identification plays a key role in ensuring security and safety in realms of household security, finance, and national defense, which makes a host of studies been proposed. 
Recent research works on human intrinsic and extrinsic properties have explored and been demonstrating promising performance. For instance, physiological characteristics~\cite{almudhahka2016human}, brain waves~\cite{zhang2018mindid}, and people's certain behavior patterns~\cite{wang2016gait}.
Most approaches are data-driven and comply with the traditional machine learning workflow. The signals and data featuring properties are first collected from various sources like wearable sensors~\cite{brutti2016online} or wireless sensing devices like Wi-Fi~\cite{xin2018freesense}. The relevant features are then identified by domain experts or extracted automatically by algorithms to represent the acquired data. Finally, the identification models are built upon machine learning or deep learning algorithms by taking these features as inputs. 
However, existing works with the workflow mentioned above are mostly static and therefore limited in handling the changing dynamics of newly observed continuous data. In real-world biometric (especially activity-based) person identification tasks, the environment semantics are naturally highly dynamic. For instance, for a gait-based identification system, the gait of a person will vary a lot in different situations. If the recognition model can not be effectively adjusted and updated accordingly, it would misidentify a person and comprise the entire identification system. Thus, the static models usually lead to accuracy decreasing in real-life usages. 

 
Except for the updating way of retraining the model after collecting enough amount of data~\cite{wen2018comparison}, some previous works explored adaptive recognition systems~\cite{pisani2019adaptive}. The assumption of these works is that there exist adequate samples to be selected before the model is updated. Nevertheless, in a real person identification system, samples are produced in a streaming manner, so that the current samples are usually not enough for retraining a static model or selecting templates. Updating the model after data accumulation is very time-consuming and unpractical. Thus, it is critical for identification systems to be self-adaptive and adjusted timely by current data, which has not been solved very well. 

To track the evolving environment more quickly and accurately with streaming data, the participation of a person with extensive knowledge and ability (called an expert), would be helpful to calibrate and justify the recognition system. The expert sits in the loop of recognition and provides reliable feedback based on the identification results, by which the recognition model is updated accordingly.
Though the idea is suitable for enabling the dynamic recognition systems with limited streaming samples, some challenges still remain: 
\begin{itemize}
\item How to effectively incorporate expert feedback in the person identification system without retraining the recognition model.
\item 
How to find the optimal updating strategy considering both recent and historical expert feedback. The strategy should balance both the immediate and future model performance at the same time.
\item 
How to improve identification performance as much as possible by keeping human efforts minimal. 
\end{itemize}

To tackle the first challenge, we treat the human-in-the-loop recognition and updating procedure as sequential interactions between the human expert and the identity recognition agent. To simplify the problem, we mainly consider vector-form streaming data such as gait information. A tree-structured model~\cite{tan2011fast} is deployed to be the base recognition model due to its explainability and scalability. To address the second and third challenges, a reinforcement learning (RL) based approach is leveraged to automatically learn the model updating strategies according to the human feedback. In the Markov Decision Process (MDP)~\cite{van2016deep}, an action can be determined by an RL algorithm based on the state and reward, which is similar to the updating strategy selection procedure based on the interaction between identification agent and human expert. Therefore, RL is appropriate to select optimal model updating strategies by reasonably combining the current identification model performance and human efforts.

In summary, we propose a novel framework which exploits \underline{\textbf{R}}einforcement \underline{\textbf{L}}earning methods to update a \underline{\textbf{T}}ree-structured \underline{\textbf{I}}dentity \underline{\textbf{R}}ecognition model (RLTIR) according to human feedback in a dynamic environment. 
The main contributions of the present work are as follows:

\begin{itemize}

\item 
We propose a self-updating interactive person identification model, namely RLTIR, which can adapt to a dynamic environment, including intra-class variations and new emerging people. 
Driven by the quality expert feedback, the recognition model can be adjusted in real time. 
Compared with existing adaptive recognition methods that need to accumulate adequate samples, our method is more suitable for improving person identification performance with streaming data. 

\item 
We develop optimal updating strategies under a reinforcement learning framework. The updating process with human expert feedback is mapped into a Markov Decision Process, which can select the most suitable updating strategy automatically at each timestamp with consideration of the current and future performance of the recognition model.

\item 
We evaluate the proposed approach on two datasets, including a local dataset and a public dataset. The experimental results illustrate that our model 
consistently outperforms the state-of-the-art methods on person identification. We also demonstrate the necessity of the feedback, robustness, and efficiency of the proposed model through a set of empirical studies. 

\end{itemize}

The remainder of this paper is organized as follows. Section~\ref{sec2} introduces literature related to this paper. Section~\ref{sec3} presents problem definition and the generic framework of RLTIR. Section~\ref{sec4} details methodology of the RLTIR method. Section~\ref{sec5} evaluates the proposed method by conducting extensive experiments on the local and public dataset and provides analysis of the experimental results. Section~\ref{sec7} gives the conclusion and proposes the future work.

\begin{figure*}[htbp]
\centering
\centerline{\includegraphics[width=0.8\textwidth]{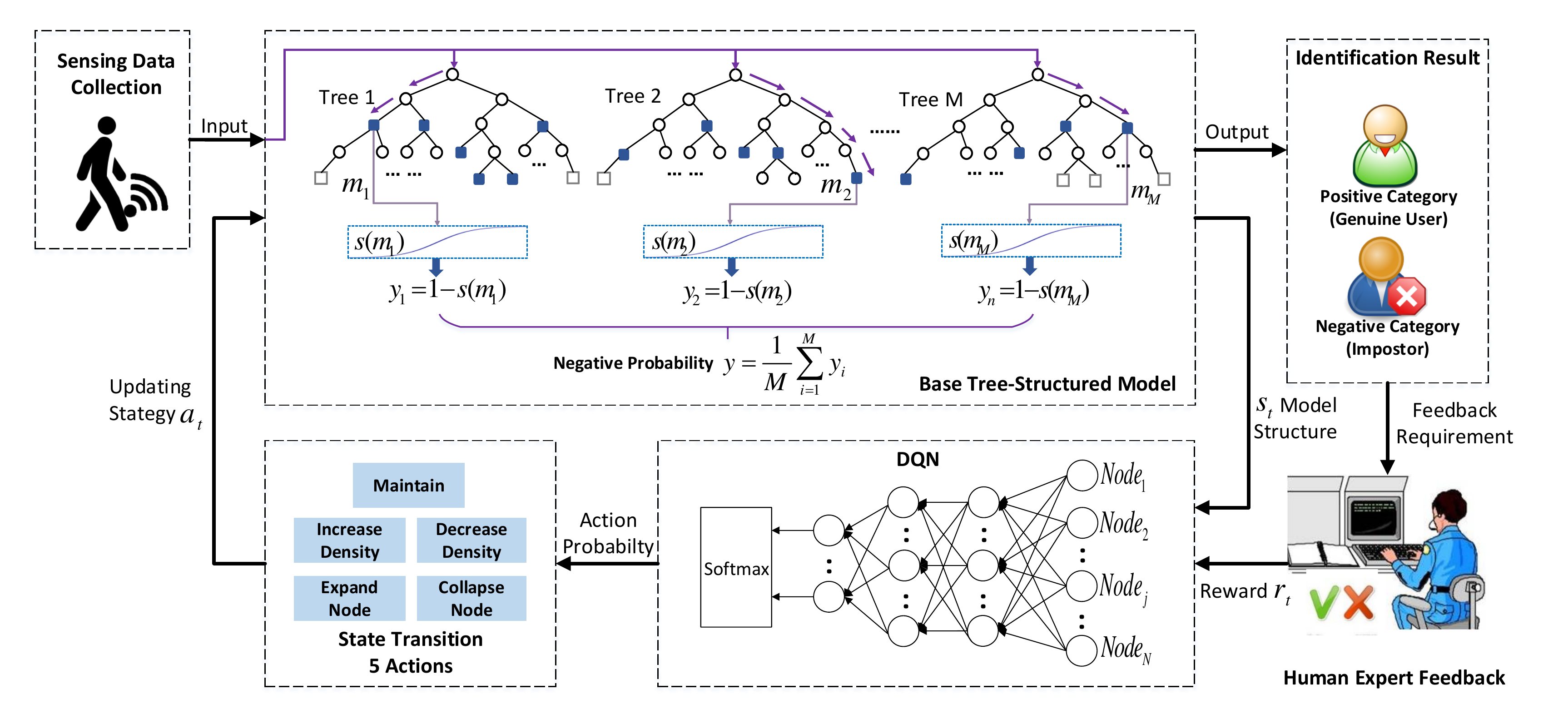}}
\caption{The overall framework of the RLTIR. For each instance of a person, an ensemble tree-structured model is used for identification. The identification result is achieved according to calculate negative probability $y$ of the instance. The model is updated according to the updating strategy learned by a reinforcement learning method in each iteration. In each step $t$, the current structures and parameters of the model can be regarded as state $s_t$. The feedback from the human expert is reward $r_t$. The DQN finds the optimal updating strategy, and the state transition selects one action $a_t$ to update $s_t$. The state transition contains five actions. Each action corresponds to an updating strategy. The updated model is leveraged to identify the next instance.}
\label{fig:framework}
\end{figure*}

\section{Related Work}\label{sec2}
Person identification has been a highly popular research topic. In recent years, biometric-based methods are usually used for identity recognition because of convenience and privacy. 
The methods utilizing fingerprints, irises, and facial features require the user to be close to the sensing device. Thus, the recent advances in device-free approaches have been extensively studied due to convenience and portability~\cite{zhang2017toward,wu2017device,shao2018bledoorguard}. Device-free approaches learn the signature of individual behaviors from the variations caused by human-environment interactions under people's behavior patterns in an obtrusive way~\cite{he2019high, wang2017risky}. 
Human activities like gait information~\cite{cao2018radar, xu2019acousticid} and keystroke dynamics~\cite{killourhy2010did} provide a convenient, low cost, and universal solution for person identification. People can be detected based on changes of the corresponding signal pattern. However, compared to systems based on face images and fingerprints with robust feature sets, signals of activities are not immutable, and there exist changes during the data collection for the same person. Different from the static models that need to be retrained, our proposed method can improve the recognition performance to promote the practicality and stability of the activity-based person identification in a better way.

To adapt to the dynamic environment, self-adaptive systems are proposed, which have been extensively studied in many works~\cite{saputri2020application,wang2019self,sun2020towards}. Self-adaptive methods can be utilized in recognition systems to improve the performance from the perspective of model structure. For example, B. Freni et al.~\cite{freni2008template} and G. Orrù et al.~\cite{orru2020novel} designed self-updating methods for template-based face recognition systems. A. Mhenni et al.~\cite{mhenni2019analysis} proposed a novel user-dependent template update strategy for keystroke recognition.
Most of methods require adequate samples before updating strategies are carried out. In a real person identification system, samples are always produced in a streaming manner, so that updating the model after data accumulation is time-consuming. To achieve the real information of the current data, we borrowed the idea from the human-in-the-loop systems~\cite{ambati2011active, zhang2018similarity}, which arranged humans to annotate samples for obtaining or updating a training set. Active learning is a classical learning paradigm leveraged in human-in-the-loop systems. 
The most typical applications are computer vision, machine translation, outlier detection~\cite{krizhevsky2017imagenet, wu2016google,liu2019generative,bhattacharjee2017active}, etc. The difference is that we set the human in both training and test process for model adaptation other than training set adaptation. In this way, our method can timely adjust the identification system only by existing and current data without accumulation of the new data.

Our work leverage reinforcement learning to bridge human guidance and the model adaptation process. Some existing studies combined traditional machine learning algorithms with reinforcement learning to obtain more interpretable and optimal models. 
Take the tree-structured model for an example, the splitting features and values were predicted for each non-leaf node orderly by employing reinforcement learning-based methods~\cite{jie2016tree, xiong2017learning, liu2018toward}. The common purpose of these works is to learn the model-building procedures and model parameters. The tree’s structure won’t be changed after training, while our model can be adapted during the test process. Sendi et al.~\cite{sendi2019new} presented a deep ensemble method based on an argumentation of combining machine learning algorithms with a multi-agent system to improve classification performance. However, the framework can not be directly employed because the interaction between the human and the identification model should be considered in our framework.  


\section{Framework Overview}\label{sec3}
The purpose of the method is to propose a new idea for timely improving activity-based person identification performance with streaming data based on the human-in-the-loop mechanism. 
To simplify the problem, it is assumed that only one person is identified at a given time.
The overall proposed RLTIR framework is shown in Fig.~\ref{fig:framework}. The workflow includes three components: base tree-structured identification model, human expert feedback, and model updating.

Tasks like sensing devices, data collection, and feature extraction are out of the scope of this article. We assume that useful features have been extracted in order to avoid interference caused by features. Most activity-based person identification systems identify persons one by one, and extract features as vector form. Considering the flexibility of the updating structure, we build a base tree-structured classification model for streaming vector-form instances of each person. For each enrolled person, the identification results are binary: positive (genuine user) and negative (impostor).
The identification result is produced according to the base identification model. The feedback on the correctness of the identification result will be given by a human expert if the result is judged as not reliable enough. The feedback will be combined with the current model structure for model updating.

For the model updating, we aim to select optimal updating strategies according to model structure changes and human expert feedback. Similarly, the core idea of the Markov Decision Process (MDP) is deciding which action to perform based on state changes in the environment and rewards obtained after performing the action. Thus, to ensure that the identity recognition model can be updated timely and reasonably by human expert feedback, we transform the sequential interactions between the human expert and the identification model as an MDP. 
Model structures and parameters are regarded as state $s$. And we set five actions for model updating, including Maintain, Increase Density, Decrease Density, Expand Node, and Collapse Node. Deep Q-learning Network (DQN)~\cite{van2016deep} is employed to select the updating strategy $a$ for the feedback in each iteration. 
The detailed updating process is demonstrated in section~\ref{sec4.4}.
The procedure is iteratively executed to timely improve the model performance without retraining the model.

\section{Methodology}\label{sec4}
This section introduces the three components of the proposed framework in detail. All the notations used in the proposed method are listed in Table~\ref{tab:notation}. 
\begin{table}[tbp]
  \caption{Notations used in this paper. The parameters that require manual settings are marked by the symbol $[*]$.}
  \vspace{-0.5cm}
  \begin{center}
  \begin{tabular}{p{2.2cm} p{6cm}}
    \toprule
    \rowcolor{mygray}
    \textbf{Parameters} & \textbf{Explanation}\\
    \midrule
    $\mathbf{x}$ & An instance need to be identified\\
    $MaxDepth^{[*]}$ & The maximum depth of each tree\\
    $MinDepth^{[*]}$ & The minimum depth of each tree\\
    $TerminalDepth^{[*]}$ & The depth of all terminal nodes in a tree, $MinDepth \le TerminalDepth \le MaxDepth$\\
    $M^{[*]}$ & The number of trees in a tree-based classifier\\
    $k$ & The chosen feature for children node splitting\\
    $\tau$ & The splitting value of the chosen feature\\
    $h$ & The depth of a node located in\\
    $m$ & The density of a node\\
    $\beta^{[*]}$ & An adaptive parameter for adaptation of $m$\\
    $v$ & The number of instances scattered in a node\\
    $y$ & The negative probability of an instance\\
    $U$ & The uncertainty of an identification result\\
    $f$ &The feedback provided by the human expert\\
    $p$ & The number of positive feedback \\
    $n$ & The number of negative feedback \\
    $\varphi^{[*]}$ & Time window size for refreshing the node\\
    $\rho^{[*]}$ & An adaptive parameter for node refreshing\\
    $s$ & The information on base tree model structure\\
    $a$ & The selected model updating strategy\\
    $r$ & The reward calculated by human feedback $f$\\
    $\gamma^{[*]}$ & The reward decay in DQN\\
    $\mathcal{D}$ & The replay memory in DQN\\
    $Q(\cdot )$ & The action-value function in DQN\\
    $L(\cdot)$ & The loss function of DQN\\
    $\theta$ & The parameters need to be learned in DQN\\
    $\alpha$ & The learning rate of DQN\\
    $\mathcal{C}^{[*]}$ & The target replace step of DQN\\
    \bottomrule
    \end{tabular}%
  \label{tab:notation}%
  \end{center}
\end{table}%

\vspace{-0.3cm}
\subsection{Base Tree-Structured Identification Model}\label{sec4.1}
Tree-structured model is a good candidate for person identification due to its explainability and scalability. As a such, the model can be effectively updated with justifications.
The identification system builds a base tree-structured classifier for each enrolled user. Each classifier contains several trees to avoid contingency. The output of each tree is the probability of the current instance belonging to the corresponding user.
Given the initial training set of an enrolled person, a random space tree is built by the feature space $\mathcal{X}\in\mathbb{R}{^d}$ that has been extracted from sensing signals. The initialization of the workspace and procedures of tree growth are similar to those in the previous work~\cite{tan2011fast}.

Each instance of a person is scattered into a specific region according to its feature vector. In the literature, the number of instances scattered in a region is related to the density of the node~\cite{wu2014rs}. 
The basic assumption is that instances of an impostor (negative category) are more possibly located in lower density regions rather than those of the genuine user (positive category). The negative probability of an instance is calculated by density from all the trees, which is used for clarifying the enrolled user and impostors. The instance with a high negative probability will be recognized as an impostor. 

A terminal node is designed for the convenience of model updating with the feedback. The depth of the terminal node is usually smaller than that of the leaf node. When an instance $\mathbf{x}$ is fed into the tree, it traverses from the root node to a terminal node, then the profiles of nodes along the path are updated. The traversing direction of the instance is determined by randomly chosen feature $k$ and splitting value $\tau$. $Node$ records characteristics and variables of the corresponding node.
A terminal node will be returned when the traversing depth $h$ equals $TerminalDepth$. In a tree, each instance corresponds to only one terminal node. 

In this part, we introduce the calculation of negative probability by returned terminal nodes. 
Let $M$ represents the number of trees in a base identification model for an enrolled user. Suppose an instance $\mathbf{x}$ locates in a terminal node of the $i$-th tree $(i \in 1, 2, \dots, M)$. The density $m_i$ of the node is
\begin{equation}\label{eq1}
	m_i = v_i \times 2^{h_i}
\end{equation}
where $v_i$ represents the number of instances scattered in the corresponding node; $h_i$ denotes the depth of the node.
The negative probability of the instance in the $i$-th tree is
\begin{equation}\label{eq2}
y_i = 1 - \textbf{s}_i(m_i)
\end{equation}
where $\textbf{s}_i(m_i)$ is the cumulative distribution function of the logistic distribution, defined as: 
\begin{equation}\label{eq3}
\textbf{s}_i(m_i; \mu_i, \sigma_i) = \frac{1}{1 + \exp \{\frac{\sqrt{3} \cdot (\mu_i-m_i)}{\pi \sigma_i} \}}
\end{equation}
where $\mu_i$ and $\sigma_i$ are the expected value and standard variance of $m_i$ for feature space $\mathcal{X}$. 
In the streaming environment, these values are incrementally calculated~\cite{welford1962note}.
Regarding the base identification model that contains $M$ trees, the final negative probability for the instance is
\begin{equation}\label{eq5}
y = \frac{1}{M}\sum_{i=0}^{M} y_i
\end{equation}
After negative probabilities of all the training instances were attained from the model, a threshold is determined. When the negative probability of a new instance is calculated, the corresponding person is identified as the genuine user if the score is lower than the threshold; Otherwise, the person is identified as an impostor.

\subsection{Human Feedback Mechanism}\label{sec4.3}
A human expert is arranged to make the person identification system timely adapt to the dynamic environment.
Although the feedback is the golden standard for improving the model performance in our method, it is unnecessary to give feedback for all the instances. Too much feedback will increase the burden of the human expert, whereas too little feedback can be not enough to improve the model. Intuitively, instances that are more likely to be misidentified require more feedback. We leverage uncertainty~\cite{shalev2012online} to evaluate the necessity of feedback for the instance. The instance with higher uncertainty is more likely given feedback by the human.

To evaluate the uncertainty of an instance, we consider the relationship between the negative probability of the instance and the number of historical feedback in the corresponding node. According to Equation~\ref{eq2}, $y_i$ is the negative probability. Suppose there has existed some feedback from the expert, the number of positive and negative feedback are $p$ and $n$. In theory, the negative probability should equal the ratio of negative feedback to all feedback, i.e., $y_i=\frac{n_i}{p_i+n_i}$. 
The uncertainty of the identification result is formulated as the difference between the expected probability $\frac{n_i}{p_i+n_i}$ and the reported probability $y_i$
\begin{equation}
u_i = |y_i-\frac{n_i}{p_i+n_i}|
\end{equation}
Considering $M$ trees in the base identification model, the final uncertainty of the instance is
\begin{equation}
U = \frac{1}{M} \sum_{i=1}^{M}u_i
\end{equation}
A threshold is set to decide whether the instance needs feedback. If $U$ is bigger than the threshold, the model will request feedback from the human expert. 

Given the feedback, the model will get access to an instance-feedback pair $\left( {\mathbf{x},f} \right)$, 
where $f \in \left\{ {0,1} \right\}$. $f = 0$ or $f = 1$ denotes that the person is recognized by the expert as positive or negative category. The identification model will be adjusted according to the feedback before getting access to future instances. 

\begin{algorithm}[tbp]
\caption{Model Updating Algorithm}
\label{alg:Update}
\begin{algorithmic}[1]
\Require $Node$ - a node in the tree, $a$ - the selected action, $\mathbf{x}$ - the instance with feedback
\Ensure An improved tree-structured model
\If{$a$ is `Maintain'}
\State Tree structure remains unchanged
\EndIf
\If{$a$ \textbf{is} `Increase Density'}
\State $Node.v = Node.v + \beta * 2^{Node.h}$
\EndIf
\If{$a$ \textbf{is} `Decrease Density'}
\State $Node.v = Node.v - \beta * 2^{Node.h}$
\EndIf
\If{$a$ \textbf{is} `Expand Node'}
\If{$Node.h < Node.MaxDepth$ }
\State $Node.left.type \leftarrow$  `terminal' 
\State $Node.right.type \leftarrow$  `terminal'
\State $Node.type \leftarrow$ `internal'
\Else 
\State $Node.v = Node.v - \beta * 2^{Node.h}$
\EndIf
\EndIf
\If{$a$ \textbf{is} `Collapse Node'}
\If{$Node.h > 1$}
\State $Node.type \leftarrow$ `internal'
\State $Node.parent.type \leftarrow$ `terminal'
\Else 
\State $Node.v = Node.v + \beta * 2^{Node.h}$
\EndIf
\EndIf
\end{algorithmic}
\end{algorithm}

\subsection{Model Updating based on Reinforcement Learning}\label{sec4.4}
The updating ways for the base tree-structured model are organized from aspects of the tree structure and the density of the terminal node. However, it is not advisable to blindly choose updating strategies. It is necessary to find an appropriate strategy to ensure that the model is optimally optimized considering both current and future performance.

We leverage reinforcement learning to update the identification model not only because the procedure of strategy selection is similar to the MDP, but also because reinforcement learning is able to balance the current and future performance. More formally, we introduce the crucial components of our deep reinforcement learning structure for the updating strategy selection:

1) \textbf{State} $s_t$: At each timestamp $t$, state $s_t = \{{Node}_1,\dots,{Node}_N\} \in \mathcal{S}$ is defined as $N$ nodes of the operational tree structure, where each node ${Node}_i$ contains a set of elements $\{h, v, p, n, k, \tau, type, flag\}$. $flag$ is a boolean value for recording if the instance has gone through the node. Other elements have been illustrated in Table~\ref{tab:notation}.

2) \textbf{Action} $a_t$: Action $a_t \in \mathcal{A}$ is defined as a strategy to update the model at timestamp $t$. We define five actions: `Maintain', `Increase/Decrease Density', `Expand/Collapse Node'. `Maintain' means $s_{t+1} = s_t$, i.e., the model remains unchanged. `Increase/Decrease Density' means that the negative probability of the terminal node will be increased or decreased. `Expand/Collapse Node' means that the terminal node is replaced by its children node or parent node. The updating processes of each action are shown in Algorithm~\ref{alg:Update}. The actions of each tree in the same classifier are independent.

\begin{algorithm}[tbp]
\caption{The RLTIR Person Identification Algorithm -- RLTIR($\mathbf{x}$, $TreeModel$)}
\label{alg:RLTIR}
\begin{algorithmic}[1]
	\Require $\mathbf{x}$ - an instance of a person, $TreeModel$ - base tree-structured model
	\Ensure $y$ - negative probability of $\mathbf{x}$
	\State Initialize replay memory $\mathcal{D}$, Initialize action-value function $Q$ with zero weights $\theta$
    \State Initialize target action-value $Q'$ with weights $\theta^- = \theta$
	\State $step \leftarrow 0$
	\State $s_t \leftarrow TreeModel.structure$
	\State $TerminalNode \leftarrow TreeModel$.Traverse($\mathbf{x}$)
	\State Calculate $y$ by Equation~\eqref{eq1}-\eqref{eq5}
	\If {$\mathbf{x}$ requires feedback} 
	\State With probability $\epsilon$ select a random strategy $a_t$; Otherwise select $\arg \mathop {\max }\limits_a Q(s_t,a_t;\theta )$
	\State Execute updating strategy $a_t$ by Algorithm~\ref{alg:Update}
	\State Observe new tree structure $s_{t+1}$ 
	\State Calculate reward $r_t$ according to feedback $f$
	\State Store ($s_t, a_t, r_t, s_{t+1}$) in Memory $\mathcal{D}$
	\State Sample random minibatch of transitions from $\mathcal{D}$
    \State Set ${l_i} = \left\{ \begin{array}{l}
    \begin{array}{*{5}{c}}
    r&{}&{\begin{array}{*{5}{c}}
    {}&{{\rm{if~identification~terminate}}}
    \end{array}}
    \end{array}\\
    r + \gamma \mathop {\max }\limits_{{a_{t + 1}}} Q({s_{t + 1}},{a_{t + 1}};\theta )\begin{array}{*{5}{c}}
    {}&{otherwise}
    \end{array}
    \end{array} \right.$
    \State Perform a stochastic gradient descent step on $({l_i} -         Q{({s_t},{a_t};\theta )^2})$
	\State step+=1
	\State $s_t \leftarrow s_{t+1}$
	\If {$step >=\mathcal{C}$}
	\State replace $\theta^-  \leftarrow  \theta$
	\EndIf
	\State Calculate $y$ using updated model with structure $s_{t+1}$ 
	\EndIf
	\State \textbf{Return} $y$
\end{algorithmic}
\end{algorithm}

3) \textbf{Reward} $r_t$: Human feedback is related to the reward $r_t$ to evaluate if the updating strategy $a_t$ is optimal. The base classifier is an ensemble model with $M$ trees. The final result is attained from results of $M$ trees. Thus, we set $r_{t, global}$ and $r_{t, regional}$ to represent the reward of the whole model and each tree. If the classifier identifies the person correctly, $r_{t, global} = +1$; Otherwise, $r_{t, global} = -1$. $r_{t, regional}$ is designed the same as $r_{t, global}$. For each tree in the classifier, $r_t = \sigma * r_{t, global} + (1 - \sigma) * r_{t, regional}$.

4) \textbf{Policy} $\pi(s_t, a_t)$: When a strategy $a_t$ is implemented on the current $s_t$, $s_{t+1}$ can be achieved. At timestamp $t$, the state transition only chooses one action following the policy $\pi(s_t, a_t)$.

\begin{table*}[htbp]
  \caption{The performance comparison with baselines and the state-of-the-art methods on different datasets for person identification.}
  \begin{center}
    \begin{tabular}{p{1.5cm}<{\centering}|p{0.7cm}<{\centering}|p{3.5cm}<{\centering}|p{1.0cm}<{\centering}|p{1.0cm}<{\centering}|p{1.5cm}<{\centering}|p{1.0cm}<{\centering}|p{1.0cm}<{\centering}|p{1.0cm}<{\centering}}
    \toprule
    \rowcolor{mygray}
    \multicolumn{1}{c|}{\textbf{Dataset}} & \textbf{Index} &\textbf{Method} & \textbf{Precision} & \textbf{Recall}& \textbf{F1-Score} & \textbf{AUC}& \textbf{FNR} & \textbf{FPR} \\
    \midrule
    \multicolumn{1}{c|}{}&{1}& {(Filippov et al. 2018)} &0.6800& 0.4139 & 0.4869 &0.6195  & 0.5861  & 0.1749 \\
    \multicolumn{1}{c|}{}&{2}&{(Wang et al. 2017)} & 0.5200 & 0.2209 & 0.2933 & 0.6018 & 0.7791 & 0.0173\\
    \multicolumn{1}{c|}{}&{3}&{(Zhang et al 2019)} & 0.7693 & 0.4573 & 0.5716 & 0.7125 & 0.5427 &0.0322\\
    \multicolumn{1}{c|}{}&{4}&{(Hejazi et al. 2017)} & 0.6252 & 0.8479 & 0.7104 &0.6508 & \textbf{0.1521} &0.5463 \\
    \multicolumn{1}{c|}{}&{5}&{(Xu et al. 2019)}  & 0.5434 & 0.6310 & 0.5728 & 0.5511 & 0.3689 &0.5289 \\
    \multicolumn{1}{c|}{Gait}&{6}&{(Liu et al. 2010)}  & 0.4650 & 0.1523 & 0.2209 & 0.5676 & 0.8477 & \textbf{0.0170}\\
    \multicolumn{1}{c|}{}&{7}&{(Hossain and Chetty 2012)}  & 0.5014 & 0.5665 & 0.5157 & 0.5147 & 0.4335 & 0.5370\\
    \multicolumn{1}{c|}{}&{8}&{(Hsu et al. 2019)} & 0.4916 & 0.5260 & 0.4969 & 0.5097 & 0.4740 & 0.5067 \\
    \multicolumn{1}{c|}{}&{9}&{(Hossain et al. 2018)} & 0.6652 & 0.7328 & 0.6833 & 0.6705 & 0.4166 & 0.3947 \\
     \multicolumn{1}{c|}{}&{10}&{(Anzar et al. 2016))} & 0.8394 & 0.8161 & 0.7007 & 0.7488 & 0.3388 & 0.1476 \\
    \multicolumn{1}{c|}{}&{11}&{Ours\_nofeed} & 0.7126 & \textbf{0.8800} & 0.7124 & 0.7219 & 0.3200 & 0.3585 \\
    \multicolumn{1}{c|}{}&{12}&{Ours (RLTIR)} & \textbf{0.8542} & 0.7785 & \textbf{0.7306} & \textbf{0.7607} & 0.4593 & 0.0869\\
    
    \midrule
    \multicolumn{1}{c|}{}&{1}& {(Filippov et al. 2018)} & 0.8286 & 0.2231 & 0.3498 & 0.76051 & 0.7769 & 0.0127\\
    \multicolumn{1}{c|}{}&{2}&{(Wang et al. 2017)} & 0.8945 & 0.8384 & 0.8064 & 0.8169 & 0.2615 & 0.0466\\
    \multicolumn{1}{c|}{}&{3}&{(Zhang et al. 2019)} & 0.9277 & 0.8796 & 0.8987 & 0.8943 & 0.1203 & 0.0310\\
    \multicolumn{1}{c|}{}&{4}&{(Hejazi et al. 2017)} & 0.6686 & 0.8784 & 0.7579 & 0.7225 & \textbf{0.1216} & 0.4333\\
    \multicolumn{1}{c|}{}&{5}&{(Xu et al. 2019)} & 0.7742 & 0.6835 & 0.8033 & 0.7417 & 0.4164 & 0.1245\\
    \multicolumn{1}{c|}{CMU}&{6}&{(Liu et al. 2010)} & 0.8813 & 0.8672 & 0.8067 & 0.8789 & 0.2328 & 0.0931\\
    \multicolumn{1}{c|}{}&{7}&{(Hossain and Chetty 2012)}  & 0.7929 & 0.8355 & 0.8089 & 0.7967 & 0.1644 & 0.2422 \\
    \multicolumn{1}{c|}{}&{8}&{(Hsu et al. 2019)} & 0.7891 & 0.7931 & 0.7881 & 0.7907 & 0.2069 & 0.2118 \\
    \multicolumn{1}{c|}{}&{9}&{(Hossain et al. 2018)} & 0.8707 & 0.8616 & 0.8274 & 0.8819 & 0.1383 & 0.0238 \\
    \multicolumn{1}{c|}{}&{10}&{(Anzar et al. 2016))} & 0.9207 & 0.8808 & 0.8906 & 0.9194 & 0.1325 & 0.0232 \\
    \multicolumn{1}{c|}{}&{11}&{Ours\_nofeed} & 0.9287 & \textbf{0.8985} & 0.9165 & 0.9174 & 0.1613 & 0.0207 \\
    \multicolumn{1}{c|}{}&{12}&{Ours (RLTIR)} & \textbf{0.9661} & 0.8726 & \textbf{0.9199} & \textbf{0.9235} & 0.1473 & \textbf{0.0149}\\
    
    \bottomrule
    \end{tabular}%
  \label{tab:result}%
  \end{center}
\end{table*}%

We employ DQN as the optimization policy $\pi(s_t, a_t)$ to update all the five Q-values (corresponding to the five candidate strategies in $a_t$) at each step.
The structure of each tree $s_t$ is stored as a matrix: the raw data represents each node of the tree, and the column data represents elements of each node.
We follow the standard assumption that delayed rewards are discounted by a factor $\gamma$ at each time step. Action-value function $Q(s_t, a_t)$ is defined as the expected return based on model structure $s_t$ and updating strategy $a_t$. The optimal action-value function $Q'(s_t, a_t)$ has the maximum expected return by the optimal policy. 
However, the state space in our framework is enormous because tree structures are diverse with different elements of nodes. Thus, we use the Bellman equation~\cite{bellman1966dynamic} to estimate $Q'$ for each state-action pair. 
We utilize deep neural networks with parameters $\theta$ as function approximator, i.e., $Q(s_t,a_t;\theta ) \approx Q'(s_t,a_t)$ . The Q-network is trained by minimizing a loss function:
\begin{equation}
L(\theta ) = {{\rm E}_{{s_t},{a_t},{r_t},{s_{t + 1}}}}[{(l - Q(s_t,a_t;\theta ))^2}],
\end{equation}
where $l={{\rm E}_{{s_{t + 1}}}}[{r_t} + \gamma \mathop {\max }\limits_{{a_{t + 1}}} Q(s_{t+1},a_{t+1};\theta)|s_t,a_t]$ is the target for the current iteration. The derivatives of loss function $L(\theta )$ with parameters $\theta$ are presented as follows:
\begin{equation}
\begin{aligned}
{\nabla _\theta }L(\theta ) = & {{\rm E}_{{s_t},{a_t},{r_t},{s_{t + 1}}}}[(r + \gamma \mathop {\max }\limits_{{a_{t + 1}}} Q({s_{t + 1}},{a_{t + 1}};{\theta ^ - }) \\
& - Q({s_t},{a_t};\theta )){\nabla _\theta }Q({s_t},{a_t};\theta )],
\end{aligned}
\end{equation}
where $\theta ^-$ are previous parameters, which are updated when optimizing the loss function.
Then, we utilize stochastic gradient descent to optimize the function efficiently.
Besides, we design the deep neural network as four layers. Except for the input layer, a simple RNN layer is employed in the first layer considering correlations and orders among nodes; The second and third layers are full-connected layers. We adopt Relu activate function in the first two layers, and use the softmax activate function in the last layer.
The detailed procedure of the RLTIR framework is shown in Algorithm~\ref{alg:RLTIR}.

\section{Experiments}\label{sec5}

\subsection{Experimental Settings}\label{sec5.1}

\begin{figure}[tbp]
\centering
\subfigure[Gait dataset]{%
\label{fig:treegait}
  \includegraphics[width=0.23\textwidth]{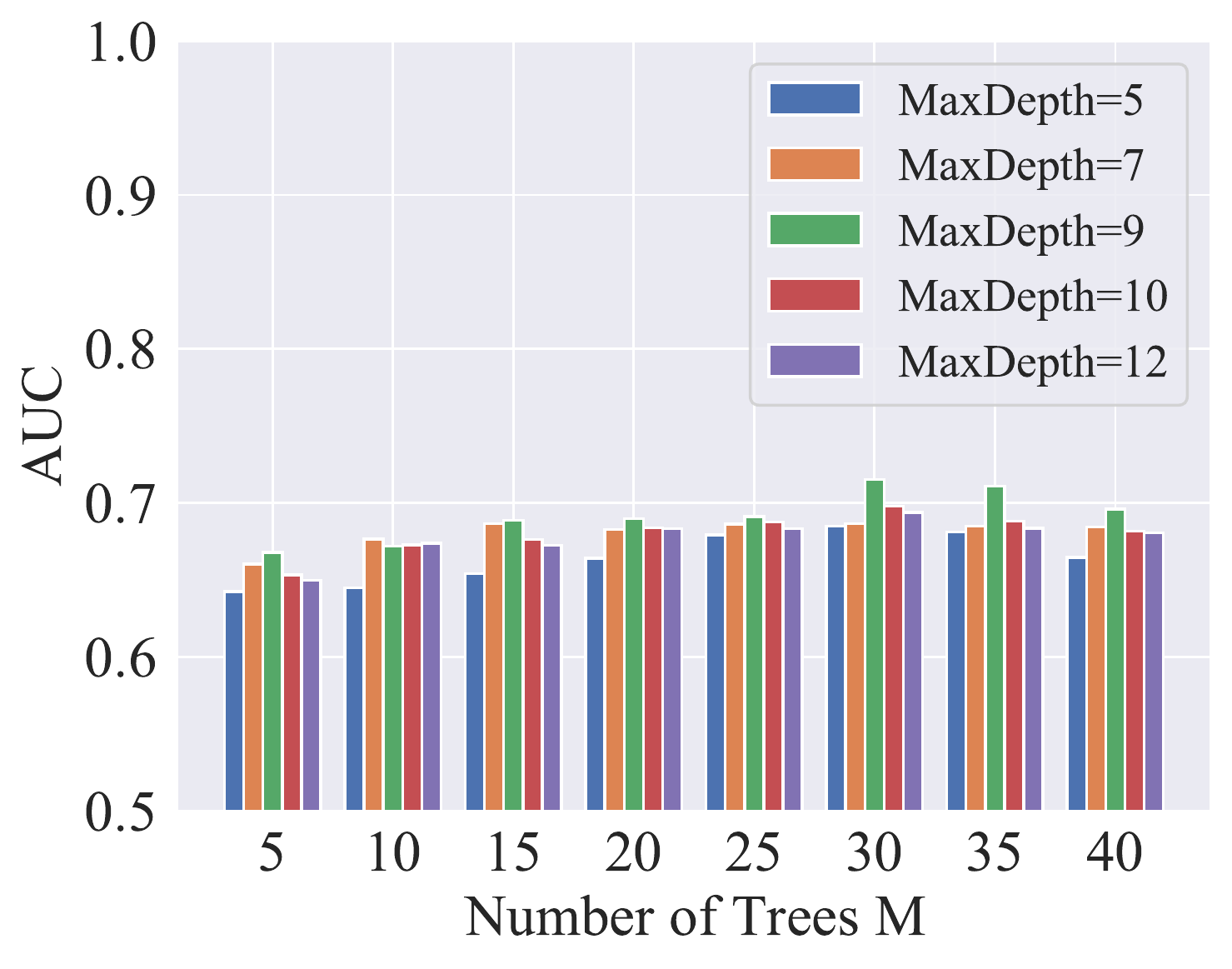}}
\subfigure[CMU dataset]{%
\label{fig:treeHAR}
\includegraphics[width=0.23\textwidth]{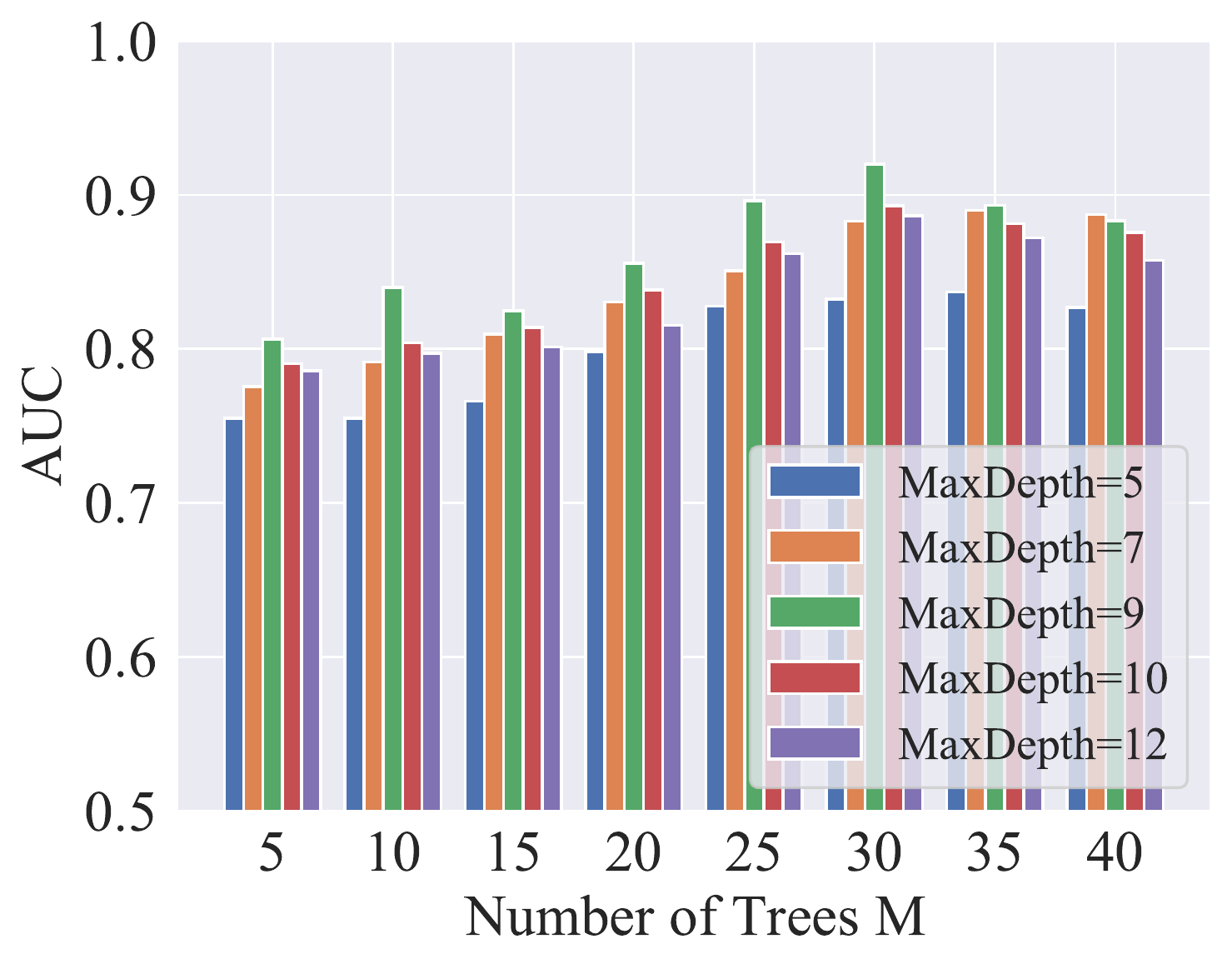}}
\caption{Area Under Curve (AUC) on different number of trees $M$ and different $MaxDepth$ of trees on two datasets.}
\label{fig:treepara}
\end{figure}

\subsubsection{Datasets}
We utilized two datasets to evaluate the RLTIR framework. The first dataset was collected from the practical person identification scenario. The second one is a public keystroke dataset named CMU~\cite{killourhy2010did} for person identification. 

The first dataset was manually attained according to the previous work~\cite{xu2019acousticid}. The work proves that the acoustic signal is effective for person identification by extracting gait patterns of the person. Therefore, we collected similar data with the same device and experimental settings in the previous work.
We recruited 50 volunteers aged between 20 and 30 years in our laboratory. 
256-dimensional features mentioned in~\cite{xu2019acousticid} were extracted from the original signal. We recorded 50 times for each volunteer in different sessions, to guarantee that gait information for each person changes temporally in our dataset. The final dataset totally contains 2,500 instances. In the beginning, only 30 people were enrolled and the rest of the persons were considered as new possible enrolled users. In the test process, when a new user appeared, we created a classifier for that user.
CMU~\cite{killourhy2010did} is a public dataset containing 51 users with 8 sessions. In each session, 50 instances were collected for each user. There are 20,400 instances in the dataset.

In the training process, for the classifier of a specific enrolled user, we randomly extracted 30\% of all available data of the user. And several data from the other enrolled ones were extracted, which accounts for 20\% of the training set of this user. For each classifier, the test set consisted of the rest 70\% data of the corresponding enrolled user and the same number of the randomly selected data from the other persons (including the new possible enrolled users). A student was arranged as a human expert. He gave feedback on whether the current instance was recognized correctly. The feedback was recorded through an interface. If the instance was recognized correctly, the expert clicked ``Yes", otherwise clicked ``No".

\subsubsection{Evaluation Metrics and Baselines}
The performance of the proposed method was evaluated by six metrics: Precision, Recall, F1-score, Area Under Curve (AUC), False Positive Rate (FPR), False Negative Rate (FNR). The baseline methods are as follows:

1. (Filippov et al. 2018)~\cite{filippov2018user} extracted features of interaction with a device's touch screen for user authentication. Isolated forest technique is employed to fit the touch pattern recognition model;

2. (Wang et al. 2017)~\cite{wang2017xrec} presented XRec which leverages user behavioral patterns of using an App. XRec employs random forest to classify device pairs;

3. (Zhang et al. 2019)~\cite{zhang2019comprehensive} investigated the sensors and features to improve the recognition accuracy of indoor activities and utilized XGBoost to recognize these activities;

4. (Hejazi et al. 2017)~\cite{hejazi2017non} found that one-class SVM can be a robust recognition algorithm for ECG biometric verification with sufficient samples;

5. (Xu et al. 2019)~\cite{xu2019acousticid} used fine-grained gait information to identify human beings. SVM classifier was utilized to be the identification model;

6. (Liu et al. 2010)~\cite{liu2010fall} developed a fall incident detection system with the help of the KNN classifier;

7. (Hossain and Chetty 2012)~\cite{hossain2012multi}  proposed an MLP-based human-identification scheme from long-range gait profiles in surveillance videos;

8. (Hsu et al. 2019)~\cite{hsu2019enabling} introduced a system that automatically collects behavior-related data and provides CNN-based algorithms for identifying activities in homes;


9. (Hossain et al. 2018)~\cite{hossain2018deactive} proposed a deep active learning-enabled activity recognition model. For fairness, we only added the most informative and meaningful sample into the training set when retrained the model was retrained.

10. (Anzar et al. 2016)~\cite{anzar2016efficient} proposed an adaptive online template-updated method using the Gaussian mixture model (GMM) to minimize classification errors and intra-class variations.

11. Ours\_nofeed is our proposed RLTIR framework without human feedback;

\begin{figure}[htbp]
\centering
\subfigure[]{%
\label{fig:weight}
  \includegraphics[width=0.23\textwidth]{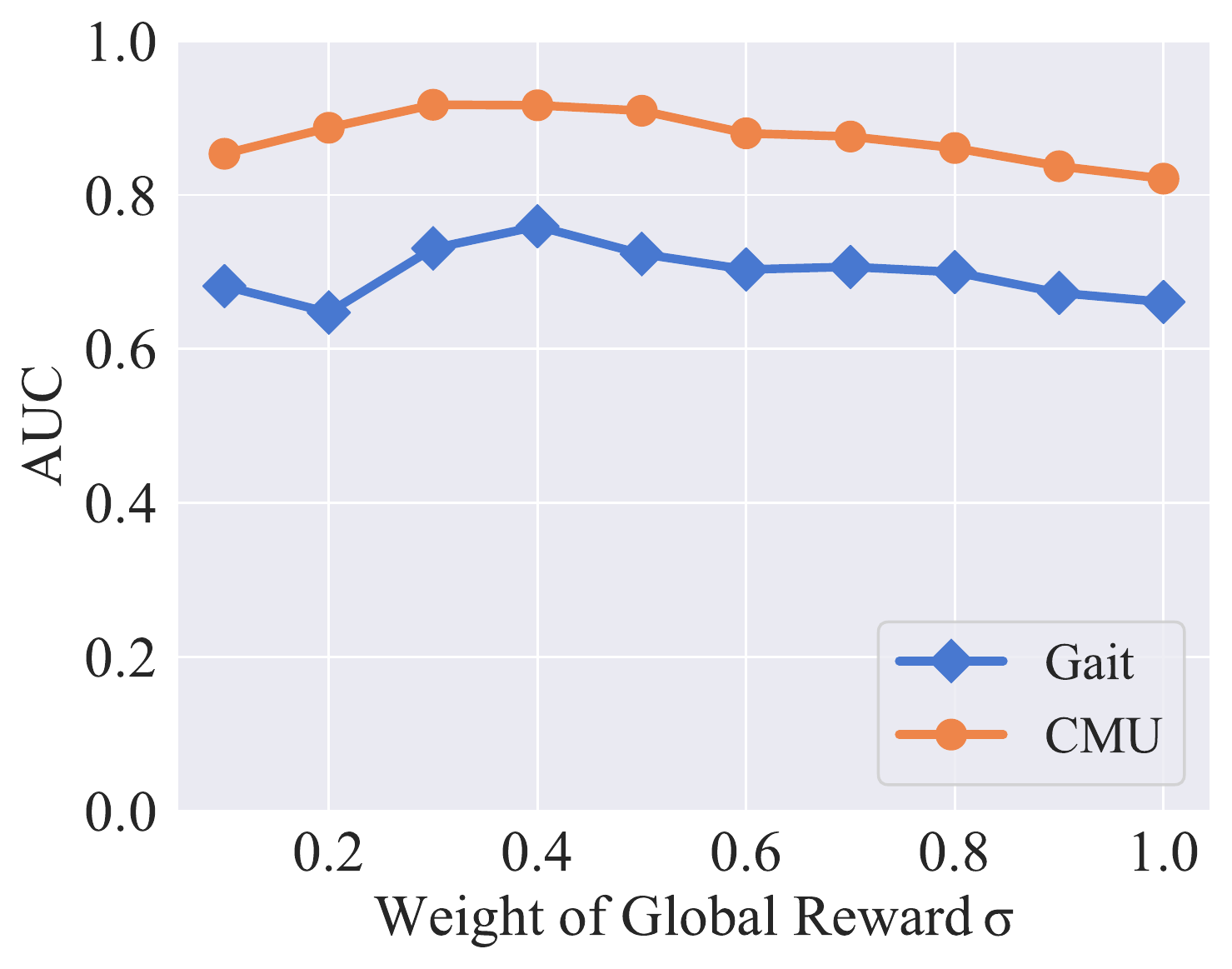}}
\subfigure[]{%
\label{fig:rewarddecay}
    \includegraphics[width=0.23\textwidth]{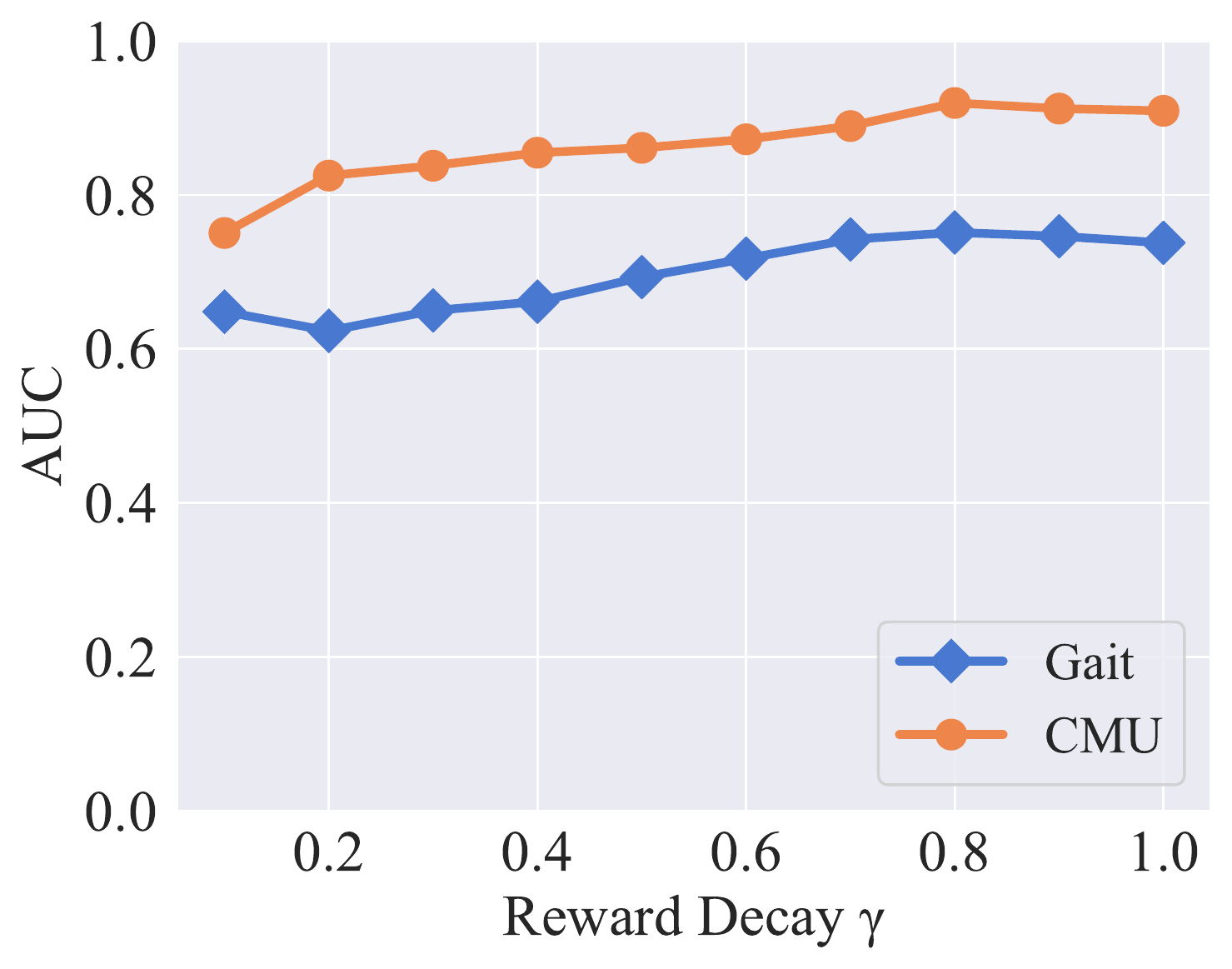}}
\subfigure[]{%
\label{fig:learningrate}
\includegraphics[width=0.23\textwidth]{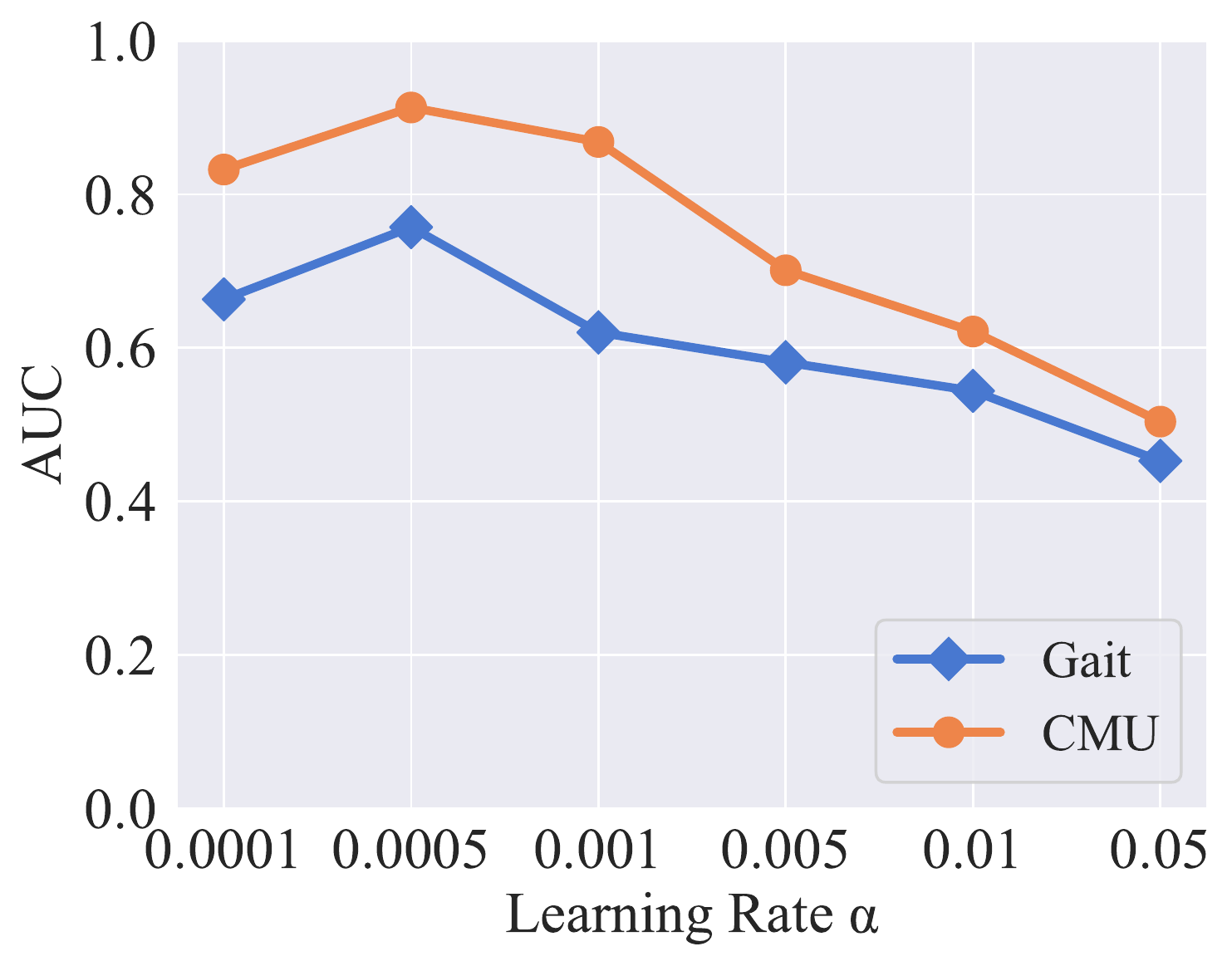}}
\subfigure[]{%
\label{fig:replacestep}
\includegraphics[width=0.23\textwidth]{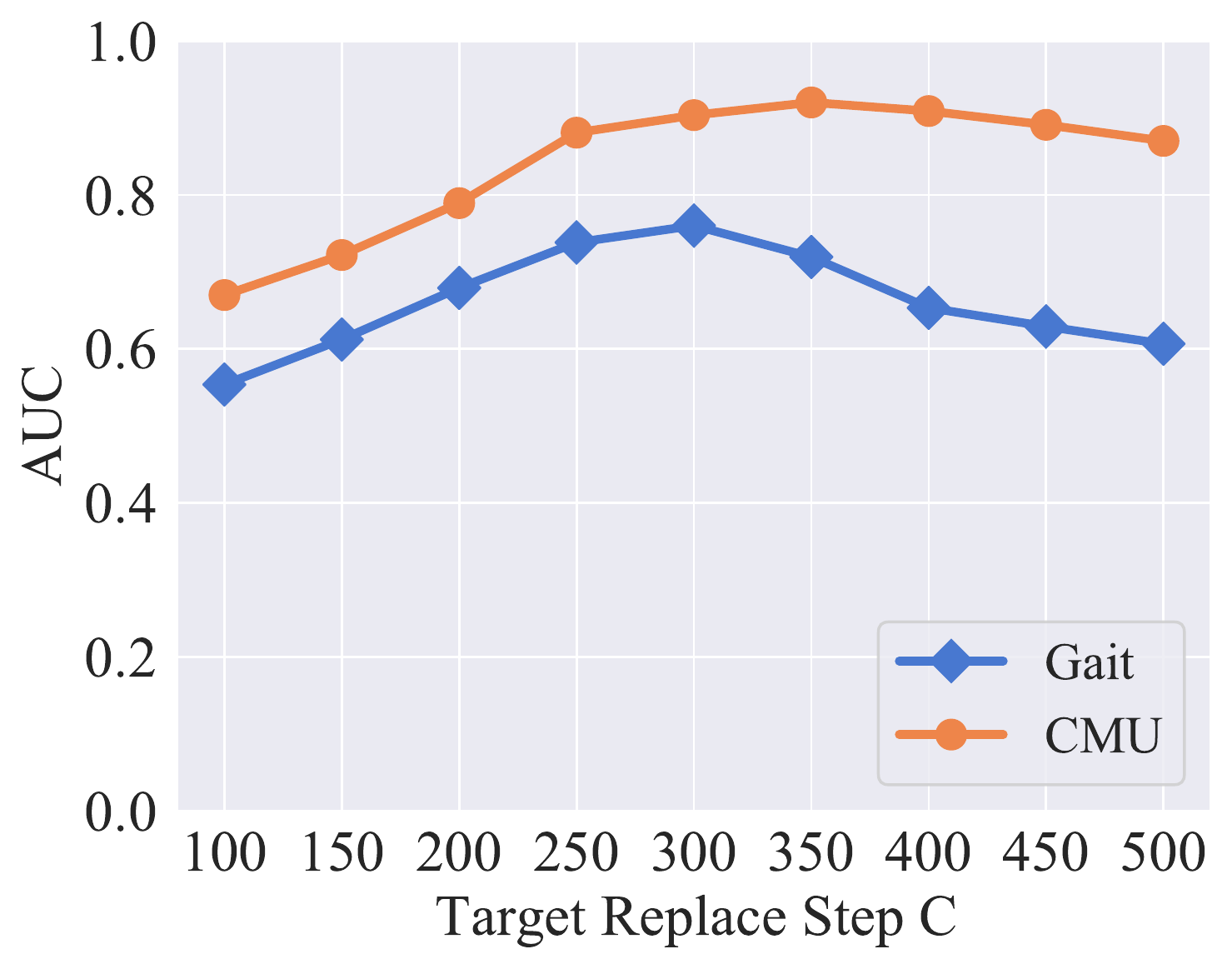}}
\subfigure[]{%
\label{fig:layer1}
\includegraphics[width=0.23\textwidth]{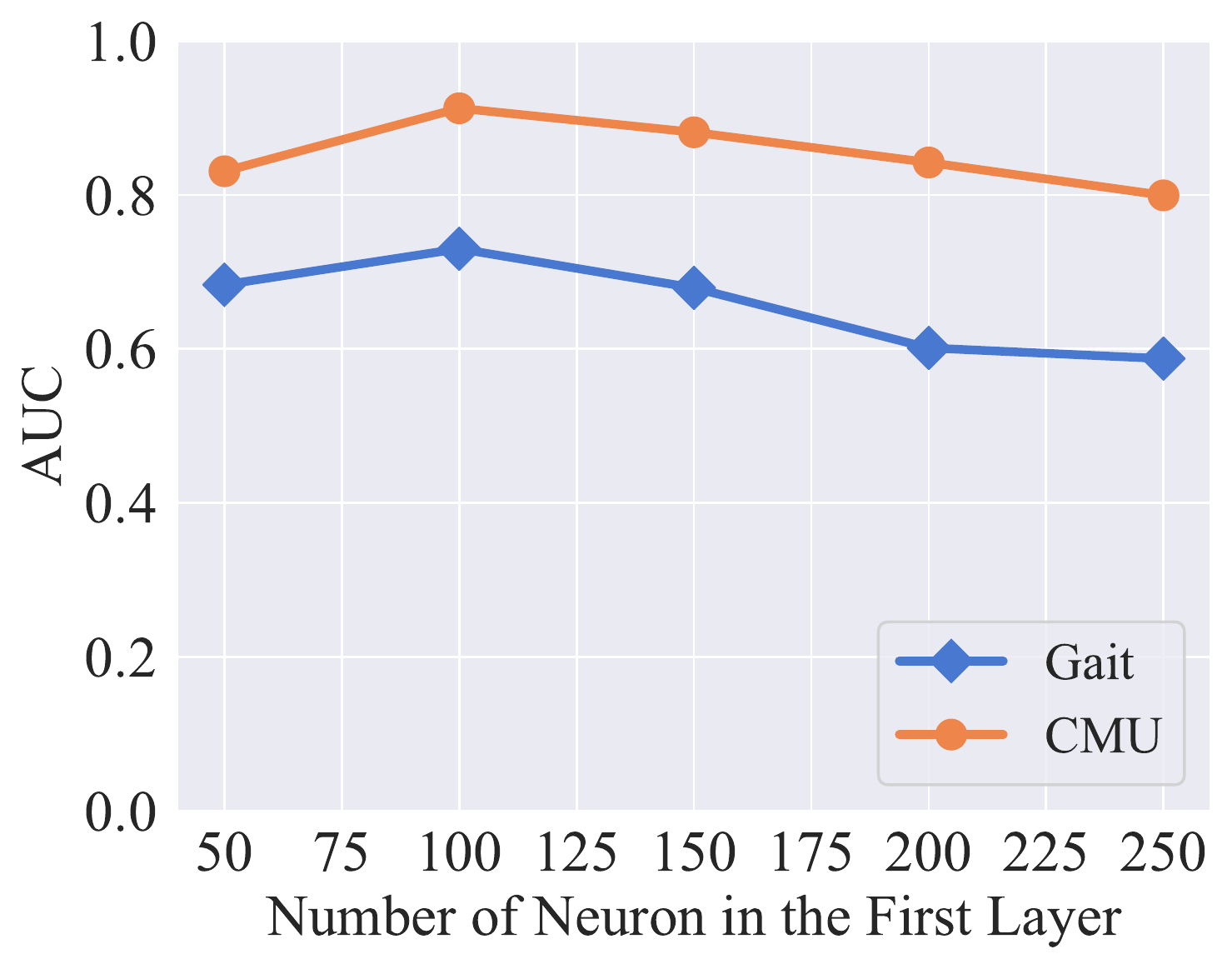}}
\subfigure[]{%
\label{fig:layer2}
\includegraphics[width=0.23\textwidth]{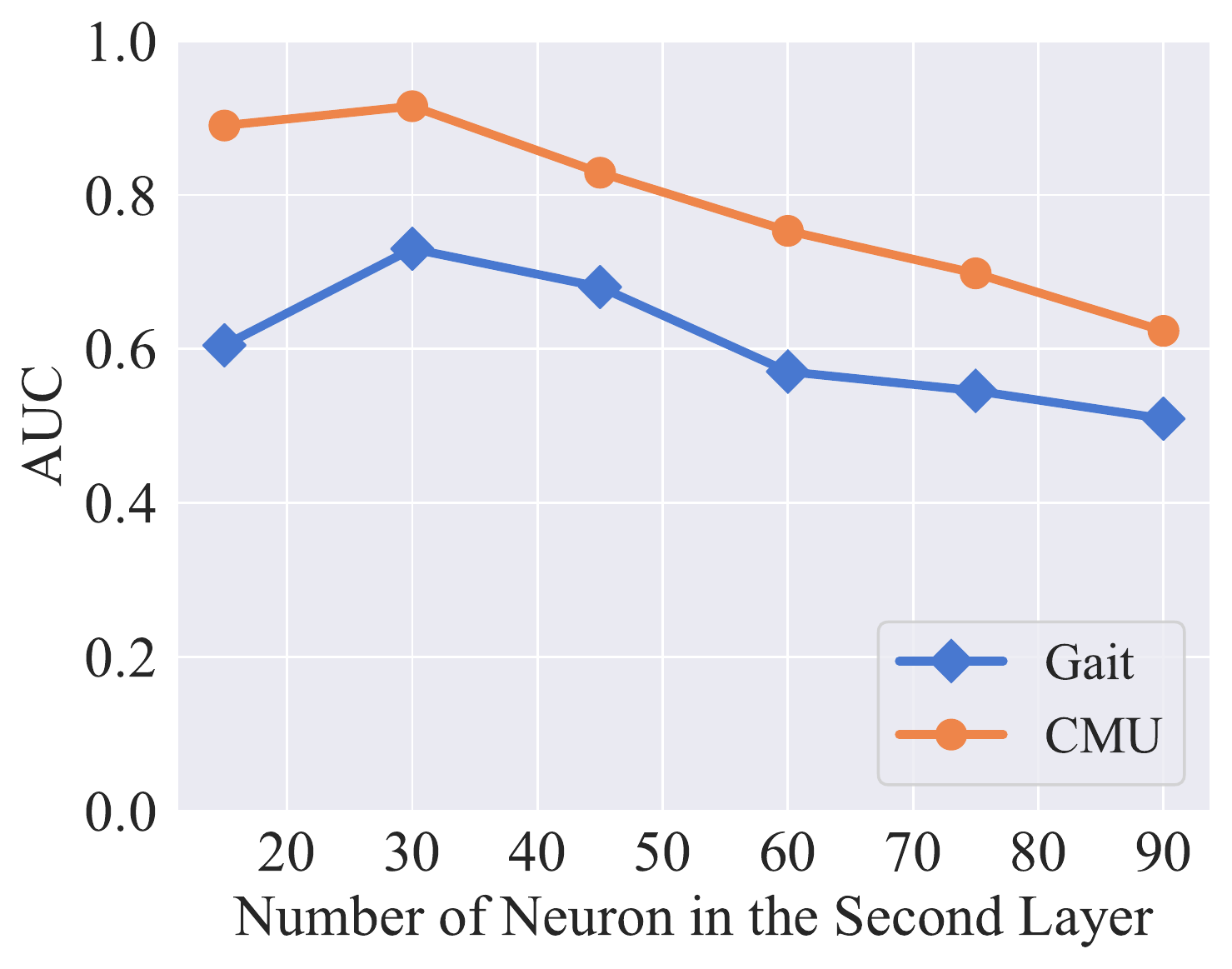}}
\caption{Area Under Curve (AUC) comparison on two datasets between (a) different weight of global reward; (b) different reward decay; (c) different learning rate; (d) different target replace step; (e) different number of neurons in the first layer; (f) different number of neurons in the second layer.}
\label{fig:DQNpara}
\end{figure}

\begin{figure}[tbp]
\centering
\subfigure[Gait Dataset]{%
\label{fig:iteration-Gait}
  \includegraphics[width=0.23\textwidth]{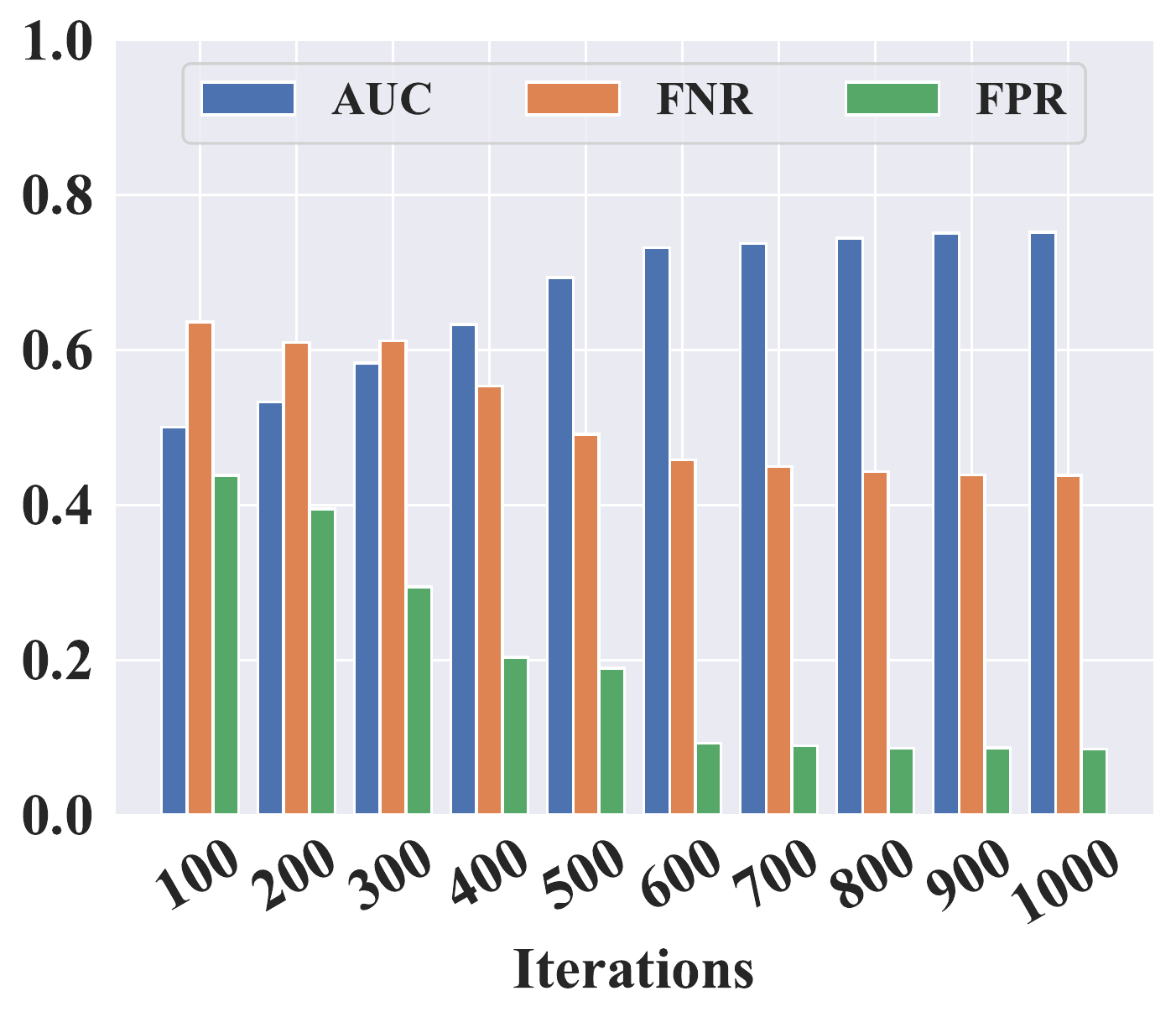}}
\subfigure[CMU Dataset]{%
\label{fig:iteration-HAR}
    \includegraphics[width=0.23\textwidth]{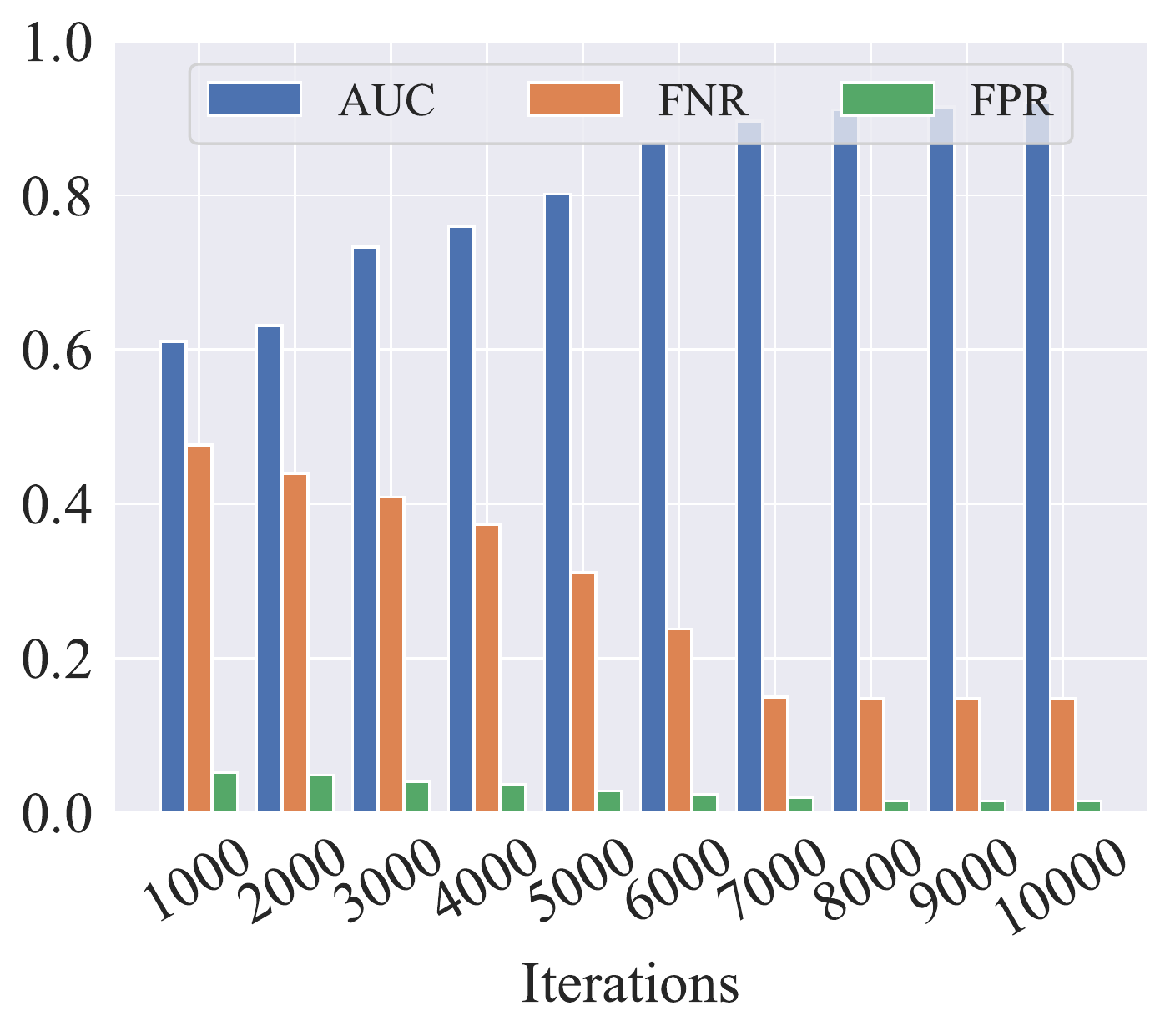}}
\caption{The change of AUC, FNR, and FPR during iterations on (a) Gait dataset and (b) CMU datset.}
\label{fig:iteration}
\end{figure}

\begin{figure*}[tbp]
\centering
\subfigure[]{%
\label{fig:feedback-proportion}
  \includegraphics[width=0.25\textwidth]{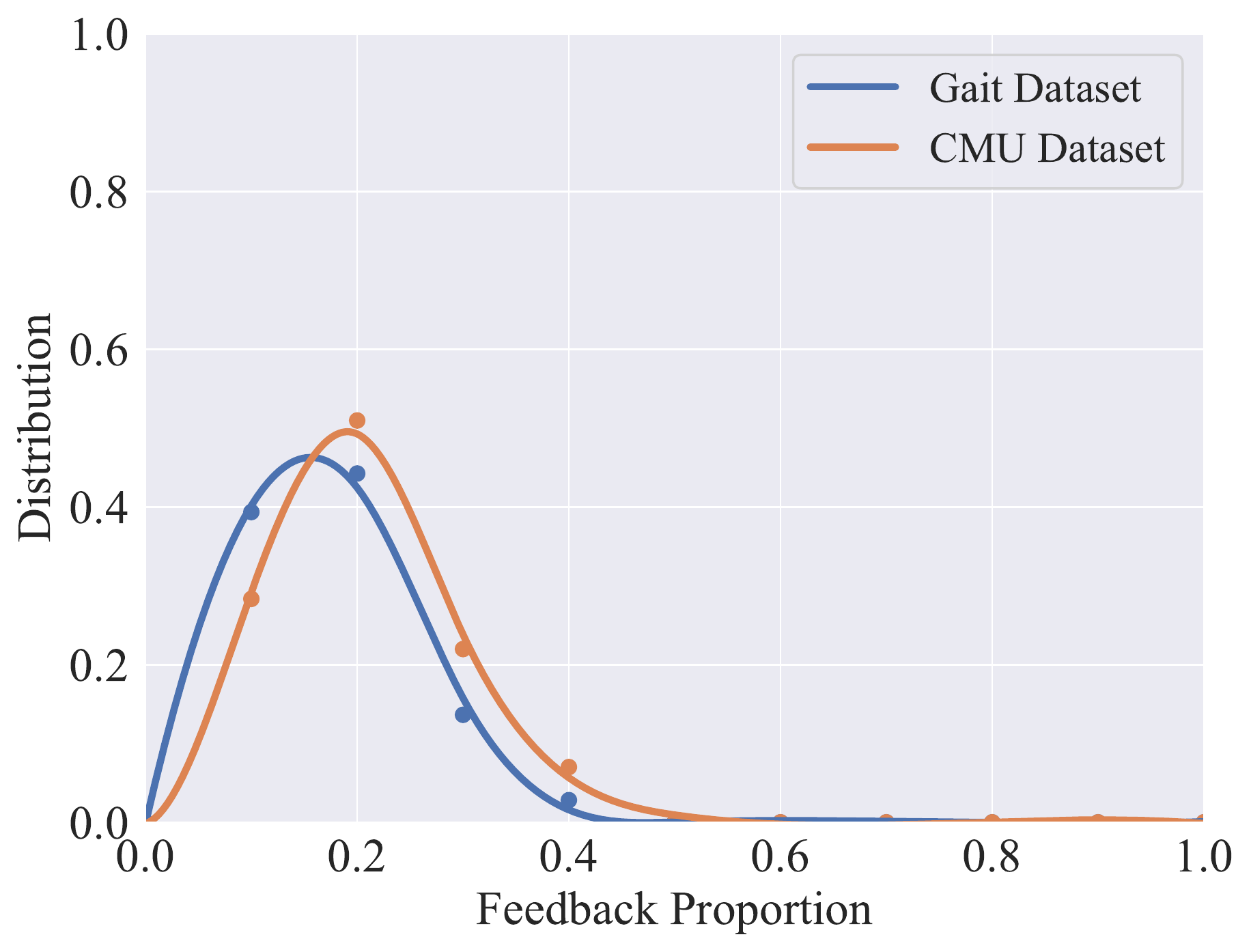}}
\subfigure[]{%
\label{fig:feedback-number-gait}
\includegraphics[width=0.25\textwidth]{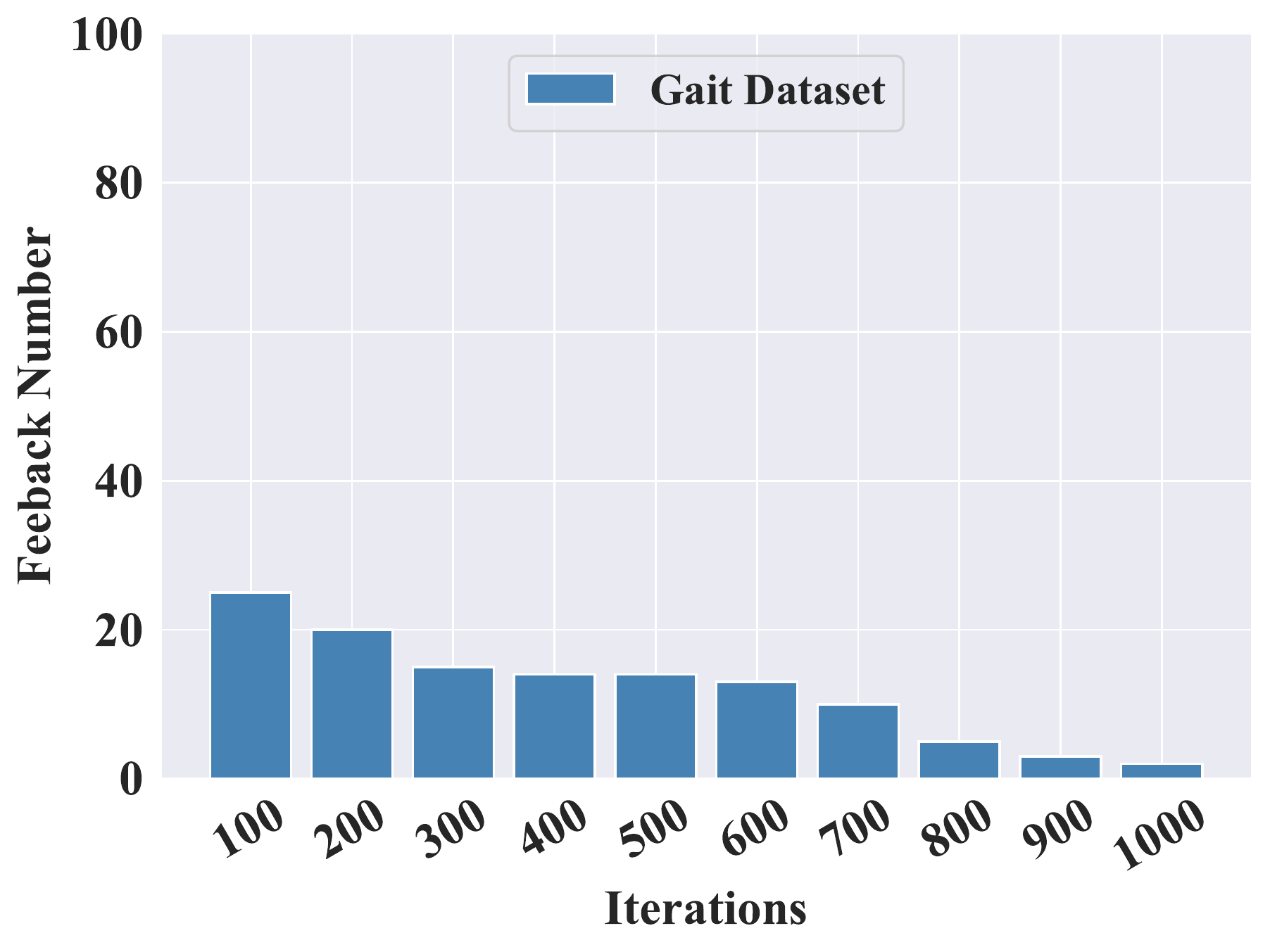}}
\subfigure[]{%
\label{fig:feedback-number-CMU}
\includegraphics[width=0.25\textwidth]{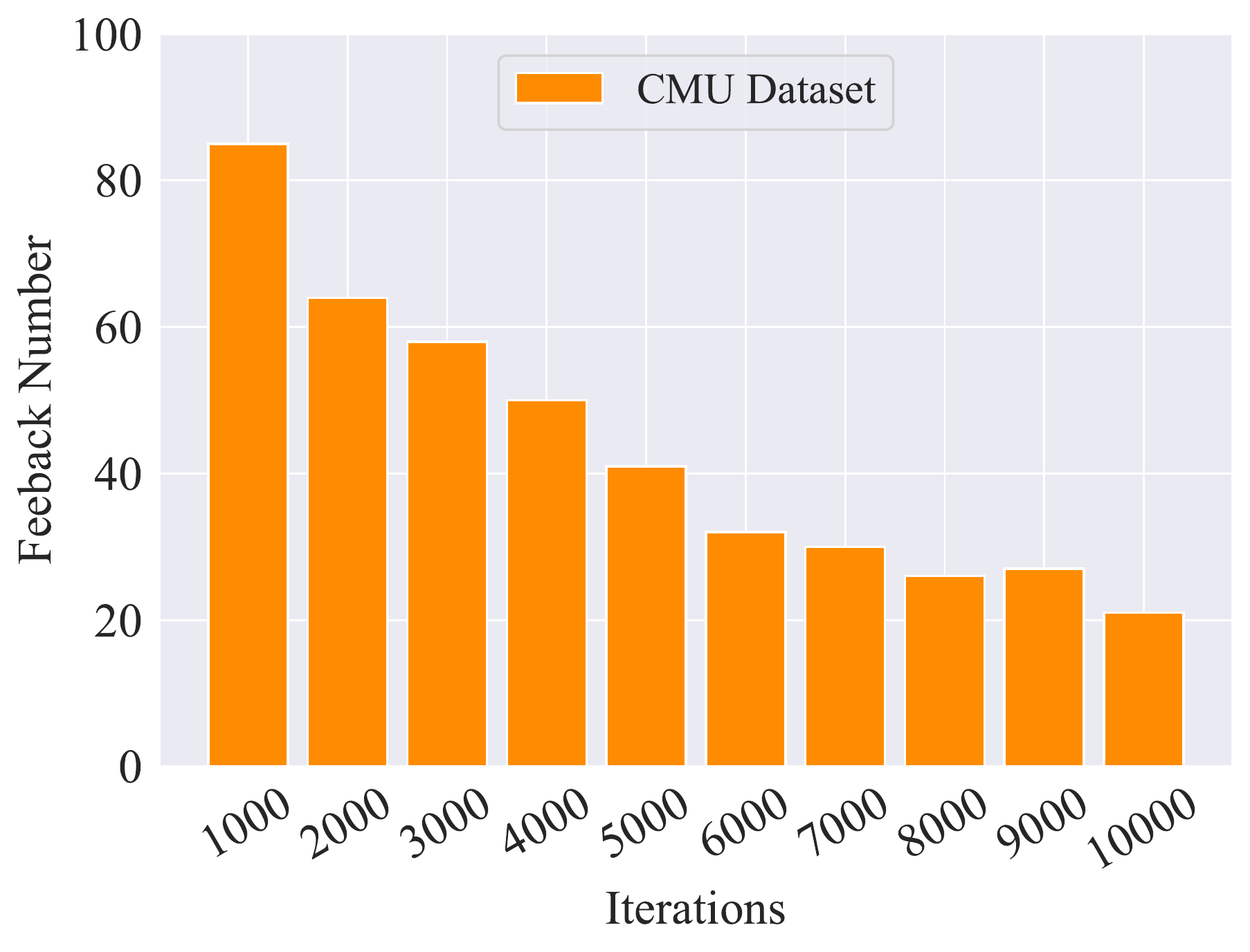}}
\caption{Analysis on the human expert feedback from two perspectives: distribution of feedback proportion on two datasets; changes of the feedback number on two datasets}
\label{fig:feedback}
\end{figure*}

\vspace{-0.2cm}
\subsection{Method Performance Evaluation}\label{sec5.2}

\subsubsection{Overall Comparison}

We compared the proposed framework with the methods mentioned above. RLTIR framework has the ability to update model structure incrementally, but most of the other methods are static. For equality, the training set of baselines contained 80\% of the instances of an enrolled user, while 30\% of the instances of the enrolled user were contained in the training set for RLTIR. We repeated the experiments 50 times and calculated average results. The overall identification results of the two datasets are shown in Table~\ref{tab:result}. It can be observed that: 
 
(1) For the Gait dataset, the proposed RLTIR framework achieves the best performance on Precision, F1-score, and AUC. Our method without feedback achieves the highest Recall. Although Hejazi's method and Liu's model attain better performances on FNR and FPR respectively, both of them perform not well on other metrics. Because the test dataset is imbalanced and the two methods identify almost all data as the same category. 
 
(2) For the CMU dataset, our method performs better than other baselines on almost all the metrics except for Recall and FNR. Anzar's model also achieves considerable performance, so it can be proved that the self-updating mechanism is helpful to improve the performances of the model.



\subsubsection{Parameter Analysis}

For the base tree-structured identification model, the maximum depth $MaxDepth$ of a tree and the number of trees $M$ affect the performance of the model. 
The experimental results on the two datasets are shown in Fig.~\ref{fig:treepara}. The model reaches the best performance when $MaxDepth$ is 9 and $M$ is 30 in all of the datasets. The performance decreases when the maximum depth of a tree is too deep, because the instance is identified partly according to historical instances scattered into terminal nodes, and a large $MaxDepth$ will cause many abundant children nodes. The performance is improved with the increasing number of trees until $M$ reaches 30.

For the updating process, we considered the weight of global reward $\sigma$, reward decay $\gamma$, learning rate $\alpha$, target replace step $\mathcal{C}$, and the number of neuron in the first two layers of DQN. 
Fig.~\ref{fig:DQNpara} shows the experimental results of these parameters. The change trends of the model on the two datasets within each parameter is roughly similar. The weight of global reward $\sigma$ and reward decay $\gamma$ do not influence the recognition performance in a wide range. The performance of our model is sensitive to four kinds of parameter settings: learning rate $\alpha$, target replace step $\mathcal{C}$, the number of neurons in the first layer, and the number of neurons in the second layer. Among all the six parameters, the learning rate is the most insensitive parameter to our model.

\subsubsection{Feedback Analysis}
To evaluate the effectiveness of the human feedback, we explored changes in the model performance during iterations on three metrics (AUC, FNR, and FPR), which are shown in Fig.~\ref{fig:iteration}. The figure illustrates that model performances on the two datasets are optimized with the increasing number of iterations. The performance achieves stability after reaching a certain number of iterations. It proves that the feedback is efficient for improving identification performance in a dynamic environment, including intra-class variations and newly enrolled users. And optimal updating strategies are gradually learned with the training of the DQN.

Although the evaluation results in Table~\ref{tab:result} have proved that the RLTIR framework outperforms, it is impractical if the frequency of feedback is high.
We hope that the model can be improved with only a few human intervention and the number of feedback will decrease with learning updating strategies constantly. 
To evaluate the burden of human experts, the overall distribution of the feedback proportion in the testing set and changes in the amount of the feedback during iterations were explored. 
Fig.~\ref{fig:feedback-proportion} reveals that the proportion of feedback from the whole testing set is almost less than 35\%. That means the expert gives feedback without taking much effort. Fig.~\ref{fig:feedback-number-gait} and Fig.~\ref{fig:feedback-number-CMU} show that feedback number decreases relatively with increasing updating iterations. After reaching a certain number of iterations, the amount of feedback tends to stabilize. It reveals that the identification performances can be improved with a limited amount of feedback.

\subsubsection{Robustness Evaluation}
\begin{figure}[tbp]
\centering
\subfigure[]{%
\label{fig:robust-ratio}
  \includegraphics[width=0.23\textwidth]{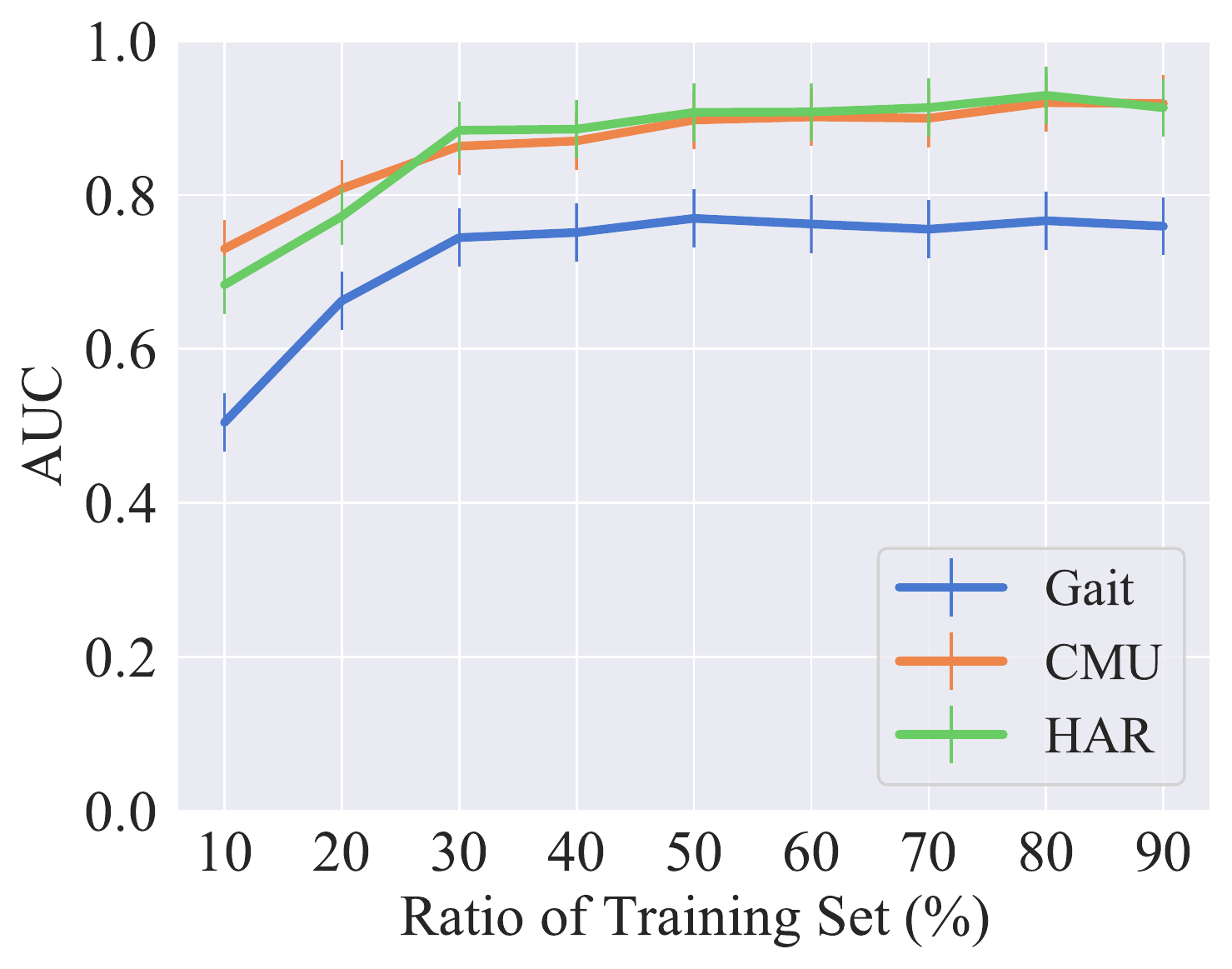}}
\subfigure[]{%
\label{fig:robust-imbalance}
\includegraphics[width=0.23\textwidth]{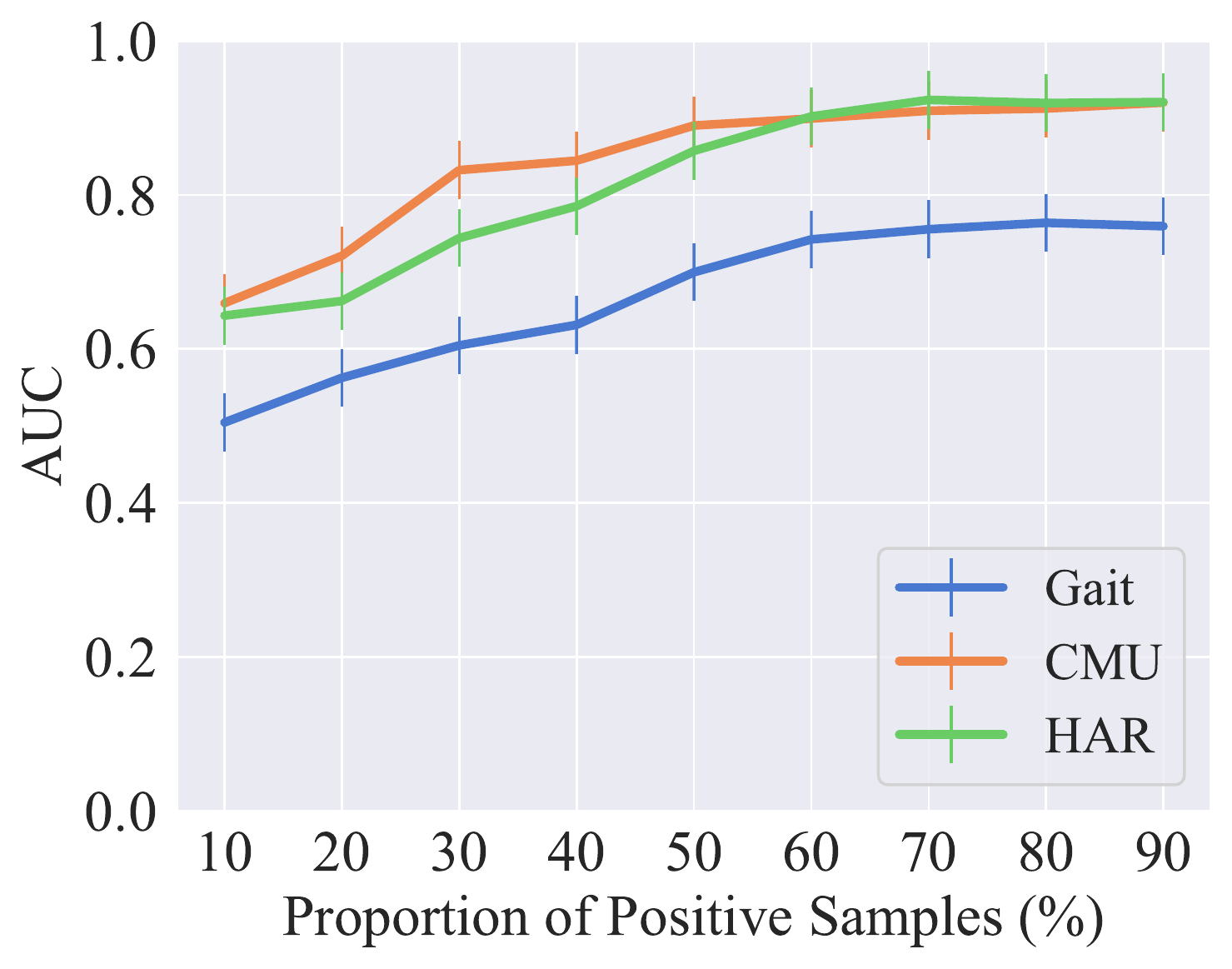}}
\caption{The change of AUC with different (a) training data size and (b) category distribution.}
\label{fig:robust}
\end{figure}

We also evaluated if the RLTIR method is robust on the size of the dataset and distribution of categories. 
we first changed the ratio of the training set to the overall dataset. Fig.~\ref{fig:robust-ratio} illustrates that the RLTIR framework is not sensitive to the size of training data as long as the ratio is higher than 30\%. For category distribution, the proportion of positive samples represents the balance between positive and negative categories.
Fig.~\ref{fig:robust-imbalance} demonstrates changes in the AUC metric with the different distribution of categories. It is obvious that the proposed method is more appropriate for imbalance category distribution, especially when there are few negative samples.

\subsubsection{Efficiency Evaluation}
\begin{figure}[tbp]
\centering
\subfigure[]{%
\label{fig:eff-train}
  \includegraphics[width=0.23\textwidth]{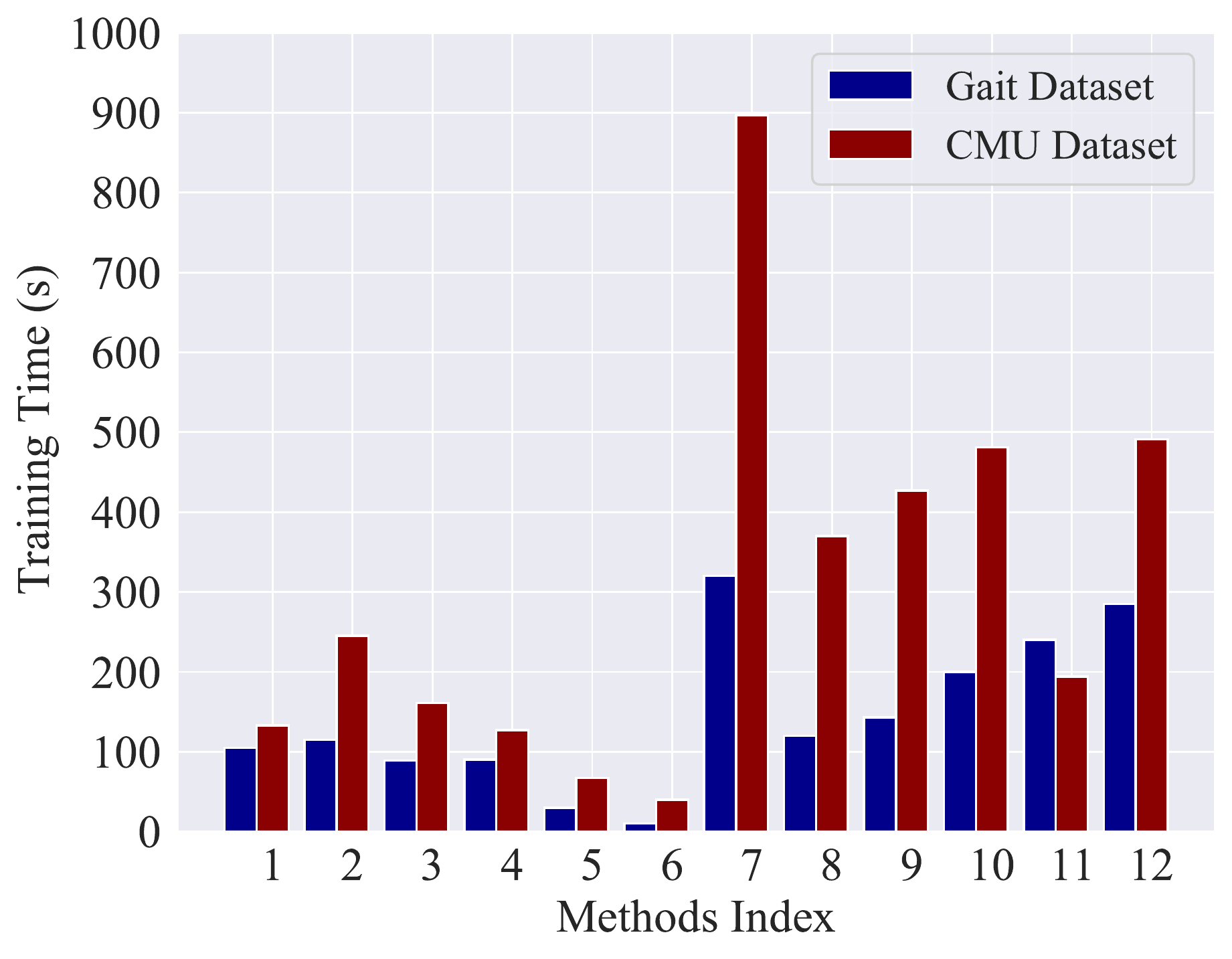}}
\subfigure[]{%
\label{fig:eff-test}
\includegraphics[width=0.23\textwidth]{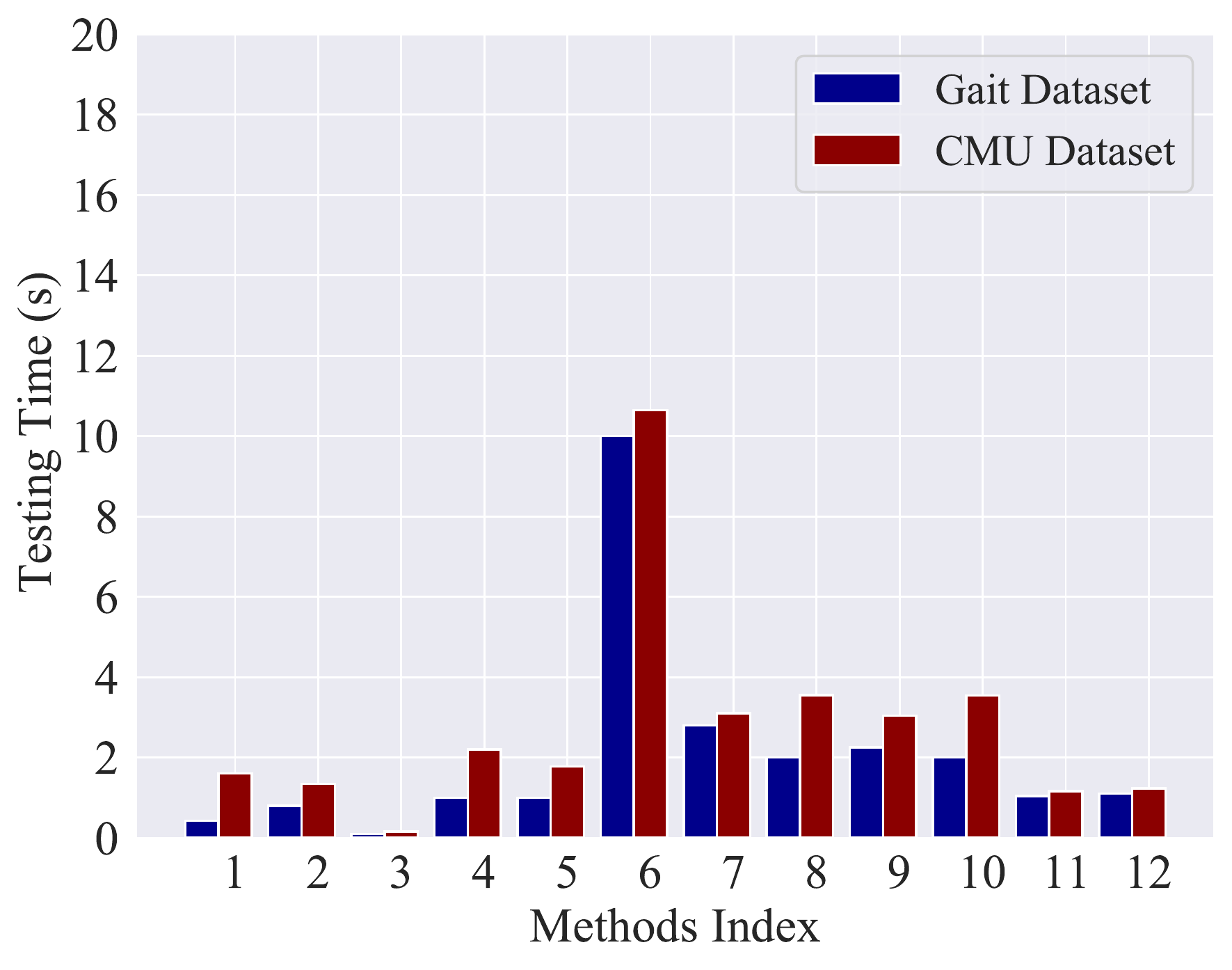}}
\caption{The efficiency on (a) training and (b) testing process for the two dataset. The index corresponds to that in Table.~\ref{tab:result}.}
\label{fig:efficiency}
\end{figure}

Efficiency refers to the required training time and identification time, including the updating process. Low efficiency will limit the suitability for practical deployment. We focused on the running time of our method and compared it with the baselines. The experiments were conducted on a computer with Intel i5-6500 3.2GHz CPU. The required training time and testing time for each method are given in Fig.~\ref{fig:eff-train} and Fig.~\ref{fig:eff-test} respectively. The x-axis denotes the index of the methods shown in Table~\ref{tab:result}). Fig.~\ref{fig:eff-train} illustrates that approaches utilizing deep neural networks take much more training time than others because these approaches contain more parameters and more complex structures. Fig.~\ref{fig:eff-test} presents that the testing time of our model is around 1 second, which is shorter than most of the other baselines. In addition, Anzar's self-updating method (No.10) costs more time on testing because it requires data accumulation and template selection. Summarily, our model is practical and efficient even considering the time cost of the updating process.

\section{Discussion}\label{sec6}
In this work, a novel method is proposed for activity-based person identification by combining human feedback with model updating process. In parallel, some limitations remain. In this section, we will discuss the limitations and potentials of the proposed method, and give the future direction of our work at the same time.
\begin{itemize}
  \item \textit{Multiple categories recognition}. In this paper, the main application scenario is recognizing if the instance belongs to the person or the current activity. In other words, recognition results only contain two categories: positive and negative. In the real world, authentication is more important than identity recognition. To address authentication problem, the model needs to be improved to fit multiple classification, so that the model could directly identify which person the instance belonging to. Simply expanding our proposed method to multiple categories recognition by building a model for each recognition entities will lead to a significant increase in testing time. Therefore, transferring the binary recognition model to a multiple recognition model considering feedback mechanism and updating strategy learning is a meaningful research issue. 
  \item \textit{Reliability of feedback}. An important component in our proposed model is feedback mechanism. The identification accuracy can be largely improved by updating the model accordingly, especially compared to existing approaches. Hence, the degree of improvement on the model depends on the reliability of feedback to some extent. In this work, we assumed the human participated in the system is an expert, so that every feedback from the human is reliable enough to improve the model. However, in the real world, the reliability of feedback from different experts is diverse. Thus, the model will be improved if it judges the reliability of the feedback and accepts the feedback with different probability. It remains an open issue for further study.
\end{itemize}
\section{Conclusion}\label{sec7}
In this paper, we propose a novel framework RLTIR for activity-based person identification by learning updating strategies incrementally based on guidance from the human expert.
The human expert gives feedback to the instance with high uncertainty. The classifier is updated according to optimal strategies learned by a reinforcement learning process. In this way, the model can be improved reasonably over time, which is a benefit for adapting dynamic environment. The proposed approach is evaluated over two datasets (a local collected dataset and a public dataset). Experimental results illustrate that our model achieves the best performance and achieves considerable robustness and efficiency with a few human efforts. 

For future work, we will transfer the binary classifier to a multi-class classifier considering feedback mechanism and updating strategy learning, so that the model will be suitable for the authentication problem. Besides, the reliability of feedback from different experts is diverse. Thus, the model will be improved if it judges the reliability of the feedback and accepts the feedback with different probability, which remains an open issue for further study.

\section*{Acknowledgment}

This work was supported in part by the National Science Fund for Distinguished Young Scholars (No. 61725205), the National Natural Science Foundation of China (No. 61960206008, No. 61772428, No. 61972319, No. 61902320), and the China Scholarship Council (award to Qingyang Li for 16 months' study abroad at the University of New South Wales).

\ifCLASSOPTIONcaptionsoff
  \newpage
\fi



\bibliographystyle{IEEEtran}
\bibliography{sample-base.bib}
%



%

\begin{IEEEbiography}[{\includegraphics[width=1in,height=1.25in,clip,keepaspectratio]{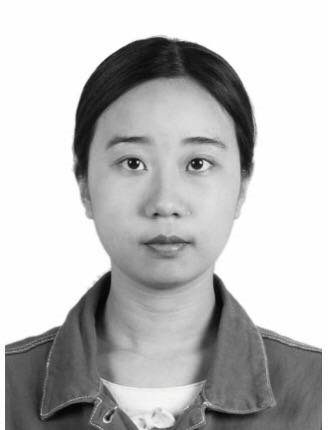}}]{Qingyang Li} 
received the bachelor’s degree from Northwestern Polytechnical University, Xi’an, China, in 2016. She is currently a Ph.D. student with the School of Computer Science, Northwestern Polytechnical University. Her research interests include ubiquitous computing, data mining, and artificial intelligence.
\end{IEEEbiography}
\vspace{-1cm}
\begin{IEEEbiography}[{\includegraphics[width=1in,height=1.25in,clip,keepaspectratio]{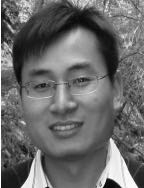}}]{Zhiwen Yu} 
received the Ph.D. degree in computer science from Northwestern Polytechnical University, Xi’an, China, in 2005. He is currently a Professor and the Dean of the School of Computer Science, Northwestern Polytechnical University, Xi’an, China. He was an Alexander Von Humboldt Fellow with Mannheim University, Germany, and a Research Fellow with Kyoto University, Kyoto, Japan. His research interests include ubiquitous computing, HCI, and mobile sensing and computing.
\end{IEEEbiography}
\vspace{-1cm}
\begin{IEEEbiography}[{\includegraphics[width=1in,height=1.25in,clip,keepaspectratio]{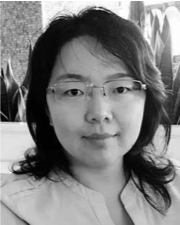}}]{Lina Yao} 
received the Ph.D. degree in computer science from the University of Adelaide, Australia, in 2014. She is currently a Lecturer of Computer Science and Engineering with the School of Computer Science and Engineering, University of New South Wales, Sydney, Australia. Her research interests include machine learning and data mining with applications to the Internet of Things, brain–computer interface, information filtering and recommending, and human activity recognition. 
\end{IEEEbiography}
\vspace{-1cm}
\begin{IEEEbiography}[{\includegraphics[width=1in,height=1.25in,clip,keepaspectratio]{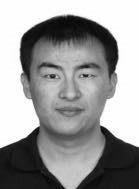}}]{Bin Guo} 
received the Ph.D. degree in computer science from Keio University, Minato, Japan, in 2009, He was a Postdoctoral Researcher with the Institut TELECOM SudParis, Essonne, France. He is currently a Professor with Northwestern Polytechnical University, Xi’an, China. His research interests include ubiquitous computing, mobile crowd sensing and computing, and HCI.
\end{IEEEbiography}




\end{document}